\documentclass[preprint,journal]{vgtc}            

\usepackage{amssymb}
\usepackage{amsmath}

\usepackage[table,xcdraw,svgnames]{xcolor}
\definecolor{TODOcolor}{HTML}{FF0000}
\definecolor{AddContentcolor}{HTML}{0000CC}
\definecolor{REVISEcolor}{HTML}{FF0000}

\newcommand{\revise}[0]{}

\onlineid{1416}

\vgtccategory{Research}

\vgtcpapertype{application/design study}

\title{VisMoDAl: Visual Analytics for Evaluating and Improving Corruption Robustness of Vision-Language Models}

\author{%
  \authororcid{Huanchen Wang}{0000-0001-9339-1941},
  Wencheng Zhang, 
  \authororcid{Zhiqiang Wang}{0000-0003-4007-7873}, 
  \authororcid{Zhicong Lu}{0000-0002-7761-6351},
  and \authororcid{Yuxin Ma}{0000-0003-0484-668X}
}

\authorfooter{
  \item H. Wang, W. Zhang, Z. Wang, and Y. Ma are with the Southern University of Science and Technology, China. Email: \{wanghc2022,~zhangwc2024\}@mail.sustech.edu.cn, wangzq\_2021@outlook.com, mayx@sustech.edu.cn. H. Wang and W. Zhang contributed equally to this work. Y. Ma is the corresponding author. H. Wang is also with the City University of Hong Kong, Hong Kong, China.
  \item Z. Lu is with the George Mason University, USA. Email: zlu6@gmu.edu.
}

\abstract{
Vision-language (VL) models have shown transformative potential across various critical domains due to their capability to comprehend multi-modal information. However, their performance frequently degrades under distribution shifts, making it crucial to assess and improve robustness against real-world data corruption encountered in practical applications.
While advancements in VL benchmark datasets and data augmentation (DA) have contributed to robustness evaluation and improvement, there remain challenges due to a lack of in-depth comprehension of model behavior as well as the need for expertise and iterative efforts to explore data patterns.
Given the achievement of visualization in explaining complex models and exploring large-scale data, understanding the impact of various data corruption on VL models aligns naturally with a visual analytics approach.
To address these challenges, we introduce VisMoDAl, a visual analytics framework designed to evaluate VL model robustness against various corruption types and identify underperformed samples to guide the development of effective DA strategies.
Grounded in the literature review and expert discussions, VisMoDAl supports multi-level analysis, ranging from examining performance under specific corruptions to task-driven inspection of model behavior and corresponding data slice.
Unlike conventional works, VisMoDAl enables users to reason about the effects of corruption on VL models, facilitating both model behavior understanding and DA strategy formulation. The utility of our system is demonstrated through case studies and quantitative evaluations focused on corruption robustness in the image captioning task.
}


\teaser{
\vspace{-1.6mm}
\setlength{\belowcaptionskip}{-0.2cm}   
  \centering
  \includegraphics[width=\linewidth]{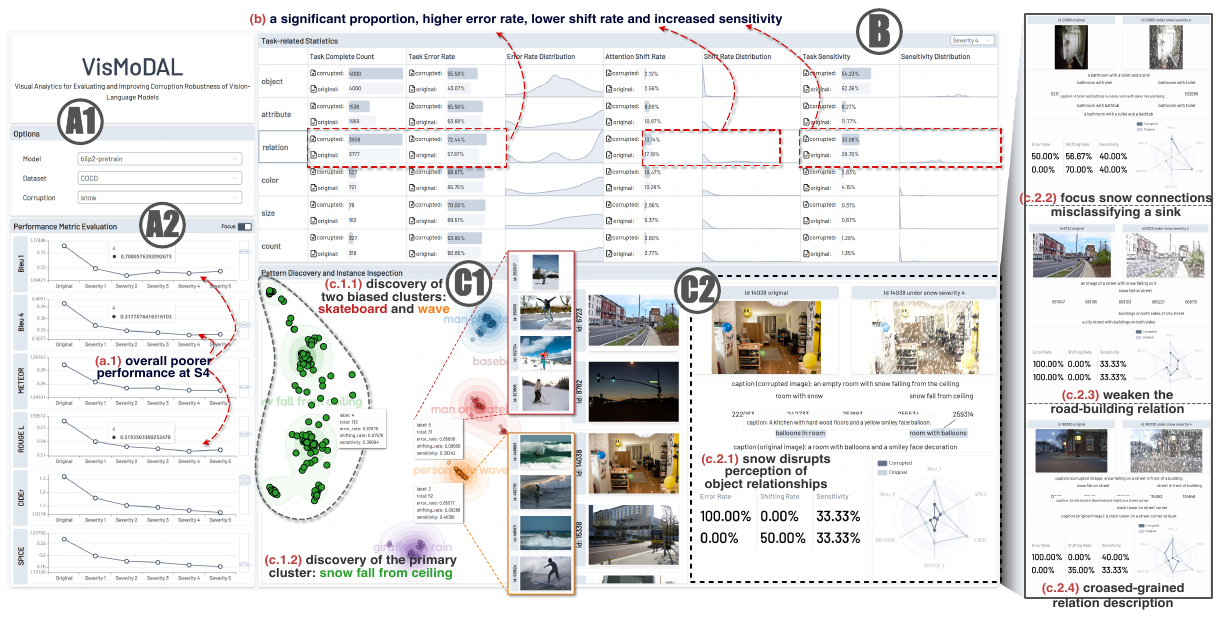}
\caption{The VisMoDAl interface comprises three key visualization components: (A) line charts presenting aggregated performance metrics for evaluation, (B) bar and distribution plots for analyzing task-driven model behavior, as well as (C) interactive scatterplot and three-layer axis view for pattern discovery and instance inspection.}
  \label{fig:teaser}
}

\graphicspath{{figs/}{figures/}{pictures/}{images/}{./}} 

\usepackage{tabu}                      
\usepackage{booktabs}                  
\usepackage{lipsum}                    
\usepackage{mwe}                       

\usepackage{mathptmx}                  

\begin{document}



 \firstsection{Introduction}
\maketitle

With the development of deep learning in recent years, research on multi-modal learning has gained momentum, which addresses the challenges of synthesizing diverse data sources with different modalities such as language and vision~\cite{Baltruvsaitis:2019:MML}. Among these endeavors, a critical point in this domain is the joint interactions between visual and textual information. Vision-Language Pre-training (VLP) exemplifies an approach designed to establish multi-modal representations from massive image-text pair datasets~\cite{Du:2022:vlp}, facilitating improved performance on downstream vision-language (VL) tasks and enabling the deployment of VL models in real-world applications, such as autonomous driving, robotic navigation, and medical diagnostics~\cite{Deng:2021:DLB, Li:2021:ICA, Bednarek:2020:ORO}.

The vulnerability of VL models to data corruption remains a major concern~\cite{Schiappa:2022:RAO, Qiu:2024:BRO}. Unlike perturbations generated by adversarial attack algorithms~\cite{Zhang:2022:TAA, Zhao:2023:OEA}, which are deliberately engineered to trigger errors, corruptions in real-world scenarios often stem from natural or technical disruptions that pollute original inputs~\cite{Hendrycks:2018:BNN, Schiappa:2022:RAO, Qiu:2024:BRO}. For example, images captured under bad weather conditions or through rapidly moving cameras may be impaired by fog, snow, or motion blur. Thus, it is important for models to counter such variations to bolster the reliability of downstream VL tasks.
Besides, attaining robustness to distribution shifts caused by data corruption is a constant challenge in deep learning. To address this issue, researchers have introduced various algorithms~\cite{Schneider:2020:IRA, Gao:2023:TBR},
metrics~\cite{Hendrycks:2018:BNN, Mintun:2021:OIB}, and benchmarks~\cite{Hendrycks:2018:BNN, Schiappa:2022:RAO, Qiu:2024:BRO} aimed at evaluating and improving model robustness. 
However, these numerical metrics-centric approaches and the complexity of VL models can provide limited semantic understanding and analysis regarding the impact of corruption on the VL model. In such scenarios, it becomes crucial to inspect the model behavior under different corruptions for effect analysis and further improvement.

Additionally, data augmentation (DA), a pivotal technique in deep learning, enhances datasets by introducing invariant properties into data~\cite{Cubuk:2019:autoaug, Cheung:2024:augCV}, thus improving model generalizability across various tasks (e.g., \textbf{object detection, relation awareness, and attribute description})~\cite{Shorten:2021:augNLP, Shorten:2019:augCV}. Notably, DA has proven effective in strengthening the robustness of VL models~\cite{Kim:2025:RobustMixGen, Zhang:2025:negaug}.
Despite the attractive result gains, traditional augmentation techniques, such as random rotation, flipping, cropping, and even the composite approaches, introduce variations in orientation and scale but are rarely effective in fine-grained scenarios and fail to capture real-world complexities~\cite{Cubuk:2019:autoaug}.
In contrast to these methods, corruptions from the real world provide a promising augmentation way to enhance the model robustness~\cite{Mintun:2021:OIB}. 
Nonetheless, the use of corruption for augmentation is often applied indiscriminately across the entire dataset. Improperly designed strategies may fail to align with task-specific requirements, resulting in suboptimal or even deteriorated model performance. Current strategies often rely on expertise to manually design or extensive tests of augmentations~\cite{Cubuk:2019:autoaug, Cheung:2024:augCV, Raileanu:2021:augRL}. 
While efforts have been made to automated approaches by finding augmentation policies for target datasets, many still face efficiency limitations~\cite{Cubuk:2019:autoaug, Li:2020:autoaug}. Moreover, establishing an interpretable connection between augmentation strategies, model behavior, and their outcomes remains a significant challenge.

\revise{The visualization community has effectively applied visual analytics approaches to comprehend model behavior and explore datasets to support model assessment and enhancement~\cite{Ma:2020:vulnerabilities, Cao:2021:ATN, Zhang:2023:sliceteller, Li:2024:va4caption, Xuan:2025:attribution, Xuan:2025:vista}.} We argue that this paradigm is applicable to evaluate and improve the robustness of VL models under corruption.
In this paper, we introduce a visual analytics framework, VisMoDAl, to inspect VL model performance and behavior across diverse corruption inputs, as well as explore those underperformed samples to further identify DA strategies.
Drawing on a literature review and expert interviews, we derived requirements and design considerations that direct the visualization and interaction design. VisMoDAl employs multi-level designs to support analyzing the effects of corruption on the VL model from three perspectives:
aggregated metrics of dataset-level performance, task-driven understanding of model behavior, and in-depth analysis of salient data pattern for augmentation.
The effectiveness of our framework is demonstrated through case studies and quantitative evaluations conducted on multi-modal dataset~\cite{Lin:2014:MCC}. In summary, our contributions include:

\begin{itemize}[leftmargin=*]
    \item A visual analytics framework, VisMoDAl, for evaluating VL model robustness against data corruption and construction of data augmentation strategies, showcased through image captioning;
    \item A suite of visual inspection and exploration methods to investigate the behavior of VL model under various data corruptions and identify the crucial patterns for data augmentation;
    \item Case studies and quantitative evaluations on a benchmark dataset to demonstrate the effectiveness of our framework.
\end{itemize}

\section{Related Work}
Our research addresses the challenges involved in analyzing the performance and behavior of the VL models under corruption.
In this section, we provide an overview of related work focusing on VL models, their robustness, and visualizations for analyzing them.

\subsection{Robustness in Vision-Language Models}
\label{sec:related-work:robustness}

\noindent\textbf{Vision-Language Models.} 
\revise{The evolution of transformer-based models~\cite{Vaswani:2017:AIA} has profoundly influenced multi-modal learning. Initially, models like BERT~\cite{Devlin:2019:BPT} and GPT~\cite{Radford:2019:LMA} demonstrated breakthroughs in language tasks, which later extended to the vision domain with Vision Transformers (ViTs)\cite{Dosovitskiy:2021:AII}, achieving remarkable results in image classification. Building on this foundation, VLP leveraged large-scale image-text datasets to learn multi-modal representations for VL tasks, utilizing transformers to integrate and reason over visual and textual information. Recent advancements, such as foundation VL models and large-scale frameworks, continue to enhance VL understanding\cite{Radford:2021:clip, Li:2022:BLI, Li:2023:blip2, Liu:2023:llava}.}

\vspace{1.2mm}\noindent\textbf{Model Robustness in Deep Learning.} Model robustness is a longstanding challenge in deep learning aimed at ensuring reliable performance in real-world scenarios. Extensive studies have been conducted on the robustness of models in visual or language domains from various aspects, including distribution shift, out-of-distribution (ood) data, and adversarial learning, as well as metrics and benchmarks for robustness evaluation~\cite{Bhojanapalli:2021:URO, Wang:2018:GAM, Vedantam:2015:CCI, Anderson:2016:spice}. Additional studies have investigated the impact of adversarial attacks~\cite{Kurakin:2017:AML, Dong:2021:HSP} and distribution shifts~\cite{Hendrycks:2018:BNN}. Furthermore, researchers have also explored techniques to improve model robustness through novel architectures and optimization methods~\cite{Madry:2018:TDL}. The multi-modal learning also recognizes the importance of robustness, particularly for VL models used in critical applications such as autonomous vehicles~\cite{Deng:2021:DLB}
, medicine~\cite{Li:2021:ICA}, and robotics~\cite{Bednarek:2020:ORO}, where robustness to corruption in real-world settings is highly required. While advancements have been made in adversarial robustness for VL models~\cite{Li:2021:AVA, Lu:2023:SGA}, corruptions often remain underexplored despite their considerable impact on model reliability.
Furthermore, current approaches for evaluating the corruption robustness of VL models~\cite{Schiappa:2022:RAO, Qiu:2024:BRO} still show some limitations. Numerous studies focus on overall performance, neglecting the effects of different corruption types; Some research provides qualitative analyses but still lacks an understanding of the impact of corruption on the model. As such, these approaches fail to offer a comprehensive explanation for the model behavior under various corruption scenarios, which can be tackled in our framework.

\vspace{1.2mm}\noindent\textbf{Data Augmentation.}  
Data augmentation serves as a fundamental technique in deep learning for enhancing the generalizability and robustness of models.
Basic DA methods, such as rotation, cropping, and flipping, are widely employed to diversify samples. Additionally, composite augmentation in uni-modal~\cite{Devries:2017:cutout, Zhang:2018:mixup, Yun:2019:cutmix} and multi-modal methods~\cite{Hao:2023:mixgen, Kim:2025:RobustMixGen, D’Incà:2024:fairaug}, designed with domain-specific knowledge, expand upon these basic techniques and have been applied to VLP and the enhancement of VL model robustness~\cite{Zhang:2025:negaug, Kim:2025:RobustMixGen}. However, such approaches may fail to address the complexities of real-world data. Real-world corruptions have been explored as potential DA methods to tackle these limitations~\cite{Mintun:2021:OIB}.
Moreover, selecting an appropriate augmentation strategy is critical, as poorly chosen policies can result in suboptimal or even degraded model performance. 
While traditional DA strategies often rely on manual design by experts or extensive experimentation~\cite{Cubuk:2019:autoaug, Cheung:2024:augCV, Raileanu:2021:augRL}, recent studies have increasingly focused on automatic policy construction~\cite{Cubuk:2019:autoaug, Li:2020:autoaug}. 
\revise{Despite advancements, some automated approaches remain inefficient and lack interpretability, hindering the understanding of relationships between data, policies, and model performance. To address this, our framework analyzes the interaction between data and VL models under corruption while introducing an interactive way to explore samples and design DA strategies to enhance robustness.}

\subsection{Visual Analytics for Deep Learning}
Visualization has emerged as an effective approach for analyzing both visual and language transformer-based models in deep learning~\cite{Yang:2024:foundation, Reif:2019:VAM, Ma:2023:VAU, Vig:2019:AMV, DeRose:2021:AFA}. Various techniques have been employed to visualize the structures and parameters of attention-based models, including projections~\cite{Li:2023:vit, Yeh:2024:AAG}, parallel coordinate plots (PCPs)~\cite{Vig:2019:AMV, Li:2023:vit}, flow maps~\cite{Shao:2023:VEF}, and heatmaps~\cite{Jaunet:2022:VXV, Zhou:2023:EVP}. 
For the language model, Shao et al.~\cite{Shao:2023:VEF} apply the attribution method to expand tree visualization to track the paths in BERT, while other studies including works by Vig~\cite{Vig:2019:AMV} and DeRose et al.~\cite{DeRose:2021:AFA} support in-depth analysis of core structures like tokens and heads across multiple layers;
For the vision model, Li et al.~\cite{Li:2023:vit} focus on head importance to help users analyze the attention patterns of ViTs. EL-VIT~\cite{Zhou:2023:EVP} aims at exploring and learning ViTs with multiple views.
With the development of VL models, visual analytics approaches have been adopted as a valuable tool for examining and understanding their behavior, aiming to improve model capability and reliability. For example, VL-InterpreT~\cite{Aflalo:2022:VIA}, offers visualizations of cross- and intra-modal components within a VL model. VisQA~\cite{Jaunet:2022:VXV} provides a solution for analyzing bias in visual question-answering models by reasoning different attention heads to predictions.

Recent visual analytics studies pay more attention to model security from model vulnerability and dataset quality. Some works identify model weaknesses under adversarial attacks~\cite{Cao:2021:ATN, Ma:2020:vulnerabilities, Deng:2025:adversal}.
While significant progress has been made in using visual analytics to interpret models and enhance unimodal robustness against adversarial attacks, the robustness of VL models under diverse corruptions remains underexplored.
\revise{Meanwhile, data-centric methods have been increasingly utilized to evaluate and improve model performance by providing insights into the data itself~\cite{Chen:2021:oodanalyzer, Chen:2022:vis4annotation, Yang:2024:labelquality, Li:2024:va4caption, Chen:2024:CVeval, Zhang:2023:sliceteller, Xuan:2025:attribution, Xuan:2025:vista}.}
MutualDetector~\cite{Chen:2022:vis4annotation} focuses on exploring annotation and improving the performance of object detector. Reweighter\cite{Yang:2024:labelquality} targets to improve the data quality by reweighting. SliceTeller~\cite{Zhang:2023:sliceteller} iteratively improves the model driven by critical data slices. Li et al.~\cite{Li:2024:va4caption} explore large-scale image datasets and evaluate the captions. \revise{Xuan et al.~\cite{Xuan:2025:attribution, Xuan:2025:vista, Xuan:2025:slim} focus on identifying and mitigating problematic data to support model validation, while also improving the quality of model outputs.}
\revise{These approaches highlight the value of analyzing model behavior from a data perspective.  Building on this foundation and addressing the above challenges, we propose a framework that leverages tasks to inspect and understand data corruption effects and identify data patterns to guide augmentation strategies for enhancing VL model robustness.}
\section{Background}
\label{background}
To illustrate our framework's applicability to the corruption robustness of VL models, this study focuses on image captioning, a representative downstream task. For this purpose, BLIP-2~\cite{Li:2023:blip2}, a foundational VL model capable of image captioning, is utilized. Subsequently, an overview of task-driven model evaluation is presented, along with background on corruption robustness.

\begin{figure}[!t]
  \centering
  \includegraphics[width=\linewidth]{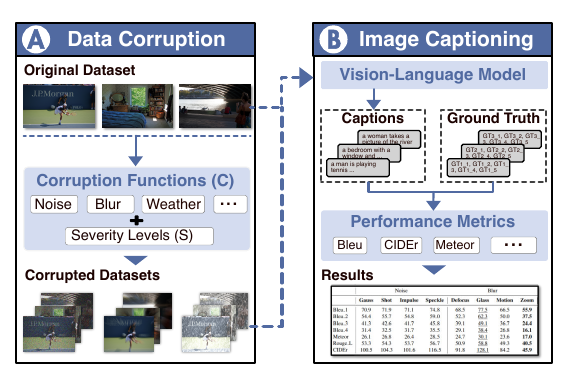}
  \caption{The pipeline for evaluating corruption robustness in image captioning comprises two primary stages: (A) the workflow for applying data corruption and (B) the subsequent process of generating image captions and evaluating the results.}
  \label{fig:background}
     \vspace{-5.6mm}
\end{figure}

\subsection{Image Captioning}
Image captioning is a fundamental task in VL understanding with widespread applications. This task bridges the gap between visual comprehension and natural language generation, demanding a model to understand the image and produce rational, semantically accurate sentences~\cite{Stefanini:2023:FST}.
The model for image captioning typically consists of two main components: a visual encoder and a text decoder.
The visual encoder extracts a representation from the input image, and the text decoder uses the representation to predict the words in the caption.
The text decoder auto-regressively predicts each word based on the generated words and the encoded image representation. 
Furthermore, as a foundation VLP method, BLIP-2~\cite{Li:2023:blip2} effectively bridges the modality gap for vision-language understanding and generation tasks, \revise{is widely utilized in image captioning and commonly serves as a baseline model in many benchmarks.}, and is thus selected for our framework.

\subsection{Task-driven VL Model Performance Evaluation}
\label{sec:task-driven-vl-model-perf-evaluation}
To enable a more granular VL model evaluation, numerous studies have adopted a task-driven approach, dividing the evaluation into multiple tasks to assess the model's capabilities across specific dimensions~\cite{Anderson:2018:task, Anderson:2016:spice}. It frequently utilizes semantic proposition categories to analyze and interpret the model's proficiency in specific capabilities, such as \textbf{object detection, relational awareness, and attribute description}, along with the subcategories of the attribute, including \textbf{color perception, counting, and understanding of size}. 
\revise{Such a framework provides insights into specific capabilities of VL models, such as their understanding of color or ability to count objects. To support task-specific evaluation, scene graph parsing (SGP) is a widely used approach for systematically analyzing the tasks performed by the VL model for each instance~\cite{Anderson:2016:spice}, as further detailed in~\autoref{view:task}.}

\subsection{Corruption Robustness} VL models encounter various corruption on input in real-world applications. To measure a model's ability to handle these corruptions, we use the term \textit{corruption robustness}. Unlike adversarial robustness, it focuses on the average performance of a model in common scenarios rather than the worst-case performance against deliberate and hard-to-distinguish perturbations.
The definition of basic corruption robustness~\cite{Hendrycks:2018:BNN} as $\mathbb{E}_{c\sim C}[\mathbb{P}_{(x,y)\sim D}(F(c(x))=y)]$ starting with a model $F$ defined with input $X$ and output $Y$. This model is trained on a dataset from the distribution $D$ and a set of corruption functions $C$, with each corruption approximating a corresponding frequency in the real world as $\mathbb{P}_C(c)$.
To evaluate corruption robustness and the impact of a specific corruption, $c$, on a model can be measured using $\mathbb{P}_{(x,y)\sim D}(F(c(x))=y)$. Thus, the performance of the VL model~\cite{Qiu:2024:BRO} in image captioning under different corruptions can integrate the specific image caption metrics to evaluate corruption robustness, as shown in~\autoref{fig:background}. Besides, to apply the corruption to the data, Hendrycks et al.~\cite{Hendrycks:2018:BNN} propose 9 corruption types, which are widely used in enhancing the model robustness~\cite{Liu:2023:corruption, Fang:2023:corruption}. Qiu et al.~\cite{Qiu:2024:BRO} further elaborate on the taxonomy for evaluating VL model robustness. The taxonomy of corruption types is used in our framework as follows: 1) Noise: Gaussian noise, Shot noise, Impulse noise, Speckle noise; 2) Blur: Defocus blur, Frosted Glass Blur, Motion blur, Zoom blur; 3) Weather: Snow, Frost, Fog, Brightness; 4) Digital: Contrast, Elastic, Pixelate, JPEG compression. Each corruption type has five levels of severity, resulting in 80 perturbation methods in total. Besides, original inputs are defined as severity level 0 in this work.

\section{Design Overview}
To analyze complex model behavior and explore large-scale data patterns under corruption for robustness evaluation and improvement, we identify key requirements based on a literature review and expert discussions. Analytical design considerations supported by the framework are subsequently derived from these established requirements.

\subsection{Requirement Analysis}
To establish a basis for the visual analysis of VL model performance under diverse corrupted inputs, we reviewed existing studies on VL model robustness analysis and evaluation~\cite{Hendrycks:2018:BNN, Hendrycks:2021:NAE, Schiappa:2022:RAO, Qiu:2024:BRO}, identifying important challenges in understanding the impact of corruptions on model behavior. In addition, we examined research on data augmentation and data quality analysis for improving VL model robustness~\cite{Kim:2025:RobustMixGen, Zhang:2025:negaug} through negative sample discovery in the design of data augmentation strategies.
We further refined our understanding of these challenges through discussions with two domain experts (E1 and E2), specializing in the VL model. E1 is a professor from the computer science department with over 15 years of experience in machine learning and computer vision, while E2 (one of the co-authors) is a PhD student in VL robust learning for six years. Both experts have published papers on the VL model in computer vision-related venues.

Based on the discussions, both experts underscored the critical role of visualization in identifying and analyzing performance degradation due to distribution shifts.
They mentioned that existing research on robustness~\cite{Schiappa:2022:RAO, Guo:2023:RTA, Qiu:2024:BRO} predominantly focuses on performance evaluation~\cite{Selvaraju:2017:GCV} and the exploration of data augmentation (DA) strategies~\cite{D’Incà:2024:fairaug, Zhang:2025:negaug}, employing basic visualization methods such as heatmaps or tabular data within notebook-based environments. These tools, however, offer limited interactivity, which constrains the analysis of model behavior and data quality, resulting in low efficiency and coarse granularity in the evaluation process. 
These limitations highlight the need for a detailed and interactive visual exploration environment, which our proposed framework aims to provide. Specifically, we focus on four key aspects of robustness evaluation and improvement under data corruption:

\begin{itemize}[leftmargin=*]
    \item \textbf{R1: Leverage Performance Metrics as a Foundation.}
    Aggregated metrics provide a high-level understanding of the VL model's performance across corruption scenarios~\cite{Stefanini:2023:FST, Qiu:2024:BRO}. These metrics serve as entry points for robustness evaluation and guide in-depth analyses.
    
    \item \textbf{R2: Incorporate Multiple Levels of Data Granularity.} 
    The analysis of robustness involves various levels of data granularity, including corruption level, task level, and instance level. Incorporating these distinct levels within a unified environment allows for a more comprehensive examination, encompassing both overall numerical indicators and fine-grained analysis of model behaviors.
    
    \item \textbf{R3: Identify and Interpret Model Behaviors.} 
    To further enhance understanding, experts highlighted the importance of not only providing semantic explanations for these behaviors but also identifying behavior changes pre- and post-corruption to derive insights into how corruption impacts the model.

    \item \textbf{R4: Extract and Analyze Data Patterns.} 
    Additionally, conducting a more in-depth exploration of data patterns is essential. Experts emphasized that extracting the data pattern with similar behavior and discrepancy under corruption, along with analyzing these instances, is crucial for understanding behavioral characteristics and guiding effective data augmentation to enhance robustness.
\end{itemize}

\subsection{Design Considerations}

Based on the aforementioned research requirements, we further outline a set of design considerations to direct our framework design.

\vspace{1.2mm}\noindent\textbf{DC1: Examine Aggregated Performance Metrics.} 
Aggregated performance metrics act as primary indicators for analysts, providing a high-level understanding of the VL model's robustness, which involves:
\begin{itemize}[leftmargin=*]
    \item Assessing the model robustness under specific corruptions; (\textbf{R1}, \textbf{R2})
    \item Examining representative trends within these metrics.  (\textbf{R1}, \textbf{R2})
\end{itemize}

\vspace{1.2mm}\noindent\textbf{DC2: Inspect Model Behaviors Across Tasks.} It is essential to associate model behaviors with task-driven semantics (e.g., \textbf{object detection, relation awareness, and attribute description}) and analyze variations at a granular level. It includes:
\begin{itemize}[leftmargin=*]
    \item  Relating model behaviors to task-specific semantic contexts; (\textbf{R3})
    \item  Investigating performance distributions across tasks and examining variations from the dataset to the instance level. (\textbf{R2}, \textbf{R3})
\end{itemize}

\vspace{1.2mm}\noindent\textbf{DC3: Uncover Data Patterns and Shared Characteristics.} It aims to uncover underlying patterns by analyzing data patterns and relationships between instances within the data patterns. It consists of:
\begin{itemize}[leftmargin=*]
    \item  Identifying the data pattern where instances exhibit similar underperformance and shared characteristics; (\textbf{R3}, \textbf{R4})
    \item  Exploring the relationships between individual samples within a pattern to understand the drivers of behavior variations. (\textbf{R3}, \textbf{R4})
\end{itemize}

\vspace{-0.1cm}
\section{Visual Analytics Framework}

Based on the identified research requirements and design considerations, we introduce VisMoDAl, a visual analytics framework designed to understand the behavior of the VL model under corruption for evaluation and improvement with three main components:

\vspace{1.2mm}\noindent\textbf{Overall Performance Analysis on Corruption (DC1).}
As the initial component of the framework, this phase offers visual representations of the VL model's performance under each corruption across all severity levels using multiple numerical metrics, in~\autoref{fig:teaser}~(A2). It also includes a comparison between the model's performance under corruption and the original (clean) condition. Such design facilitates a high-level overview of the model's robustness against specific corruptions, allowing users to identify overarching trends and recognize significant patterns within the metrics. These insights serve as a foundation for more detailed analyses in subsequent components.

\vspace{1.2mm}\noindent\textbf{Visual Inspection of Model Behavior via Tasks (DC2).}
Based on the foundational understanding of performance changes under corruption, this component provides semantic reasoning about the impact of corruption on model behavior by introducing multiple tasks as intermediaries,~\autoref{fig:teaser}~(B). This area allows users to examine the variation in the model's ability on each task and perform comparative analysis.

\vspace{1.2mm}\noindent\textbf{In-depth Visual Exploration of Data Patterns (DC3).}
Analysts are allowed to investigate the concrete behavioral characteristics of samples contributing to the results under the specific task. By employing the slice discovery method, samples with similar behaviors but reduced performance under corruption are grouped. This panel supports the identification of underlying data patterns and enables instance-level exploration and analysis, aiding in the selection of samples for targeted data augmentation, in~\autoref{fig:teaser}~(C1-2).

\subsection{Overall Performance Analysis on Corruption}\label{view:performance}
Statistical metrics offer a comprehensive evaluation of VL model performance across various corruption types and severity levels (\textbf{R1, DC1}). Furthermore, the metrics act as a reference to guide further analysis.

\vspace{1.2mm}\noindent\textbf{Performance Evaluation Metrics.} 
Drawing from established literature on the performance evaluation of VL models in image captioning~\cite{Papineni:2002:bleu, Wang:2018:GAM, Vedantam:2015:CCI, Anderson:2016:spice, Lin:2004:rouge, Banerjee:2005:meteor}, the following five standard measures, widely adopted in the evaluation of VL models, have been selected to assess robustness against corruption. These measures assess multiple aspects of the VL model's output simultaneously.
BLEU~\cite{Papineni:2002:bleu} evaluates linguistic correctness by measuring n-gram precision, focusing on word choice accuracy for $n=1$ and fluency for $n=2-4$.
METEOR~\cite{Banerjee:2005:meteor} emphasizes recall by matching unigrams between model outputs and ground truths (GTs), considering exact matches, stemming, and semantic similarity.
ROUGE\_L~\cite{Lin:2004:rouge} measures structural similarity by identifying the longest common subsequence of tokens between candidates and references, allowing for gaps.
CIDEr~\cite{Hessel:2021:cider} assesses consensus using cosine similarity of TF-IDF weighted n-grams, balancing precision and recall. SPICE~\cite{Anderson:2016:spice} focuses on semantic content by matching tuples from scene graphs of candidates and references, prioritizing meaning over fluency.
Higher values across all these metrics indicate a stronger correspondence between the model's output and the GT. When aggregated at the dataset level, these metrics not only provide a comprehensive assessment of the VL model's performance under different corruption scenarios but also enable robustness evaluation by comparing performance on corrupted inputs to that on clean inputs.

Shown in the~\autoref{fig:teaser}~(A2), this view is designed to visualize overall performance changes of standard metrics as severity levels increase under a given corruption. It consists of six line charts, each representing a distinct performance metric: BLEU1, BLEU4, METEOR, ROUGE\_L, CIDEr, and SPICE.
When a corruption type is selected, metric values are plotted on the line charts. The horizontal axis of each chart represents the severity levels of corruption, ranging from 0 (original input) to 5 (highest severity), while the vertical axis is scaled dynamically to accommodate the maximum value within the visible range of the chart. This visualization highlights the performance trends of the VL model as it changes from original to corrupted inputs with higher severity levels. It offers a comprehensive understanding of the model's robustness by analyzing trends across multiple metrics at the dataset level. When a significant trend is identified in the overview, such as a corruption type with notably lower performance metrics, the analyst can investigate the model’s behavior in detail by selecting the corresponding severity level on the performance line chart.

\subsection{Visual Inspection of Model Behavior via Tasks}\label{view:task}
Decomposing the model’s performance across multiple tasks provides a more detailed semantic understanding of the VL model's behavior, beyond aggregate metrics under specific corruption types and severity levels~(\textbf{R2, R3, DC2}). This component provides a task-driven comparison and interpretation of how the VL model's abilities vary under the selected corruption scenarios and in the original condition~(\autoref{fig:teaser}~(B)).
Building on the trend identified in the first component, the analyst delves into a more detailed examination of the model's behavior.
The task-specific behavior of the VL model is analyzed through a two-step process and visualized using bar charts and distribution plots based on task-related metrics. In alignment with \textbf{R3}, which emphasizes the need of comparing behavior changes before and after corruption to evaluate their impact on the model's robustness, the comparisons are presented across multiple dimensions in a juxtaposed manner, including the count of task-relevant samples, error rate, attention-shifting rate, and sensitivity to each task of the VL model.

\begin{figure}[!t]
  \centering
  \includegraphics[width=\linewidth]{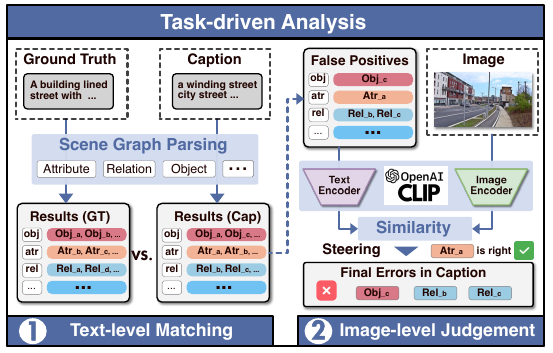}
  \caption{The pipeline of task-driven analysis of each instance.}
  \label{fig:task}
 \vspace{-6.8mm}
\end{figure}

\vspace{1.2mm}\noindent\textbf{Task-driven Model Behavior Analysis.} 
To facilitate a deeper understanding of model behavior, we incorporate tasks as a medium to establish context for analysis, the pipeline illustrated in~\autoref{fig:task}.

\vspace{1.1mm}\noindent\textit{Text-level Matching}: 
\revise{SGP is applied to examine the tasks performed by the VL model for each instance. A scene graph (SG), widely recognized for its ability to represent semantic descriptions, provides a graph-based structure where a set of \textbf{objects} classes, their \textbf{attributes} ($A$), and \textbf{relation types} ($R$) between them are constructed. Besides, $\textbf{color, count, size} \in A$ are specific attributes of objects, representing their color, quantity, and scale. Given the VL model output ($v$), the scene graph can be formally defined as a tuple~\cite{Anderson:2016:spice, Li:2023:FACTUAL}:
\vspace{-0.5mm}
\begin{equation}
    G(v) = \langle O(v), E(v) , D(v)\rangle,
\end{equation}
where $O(v) = \{ o_1, \dots, o_n \}$ represents the set of objects, $E(v) \subseteq O(v) \times R \times O(v)$ is set of edges describing the relations between objects, and $D \subseteq O(v) \times A$, which defines the set of attributes linked to the objects. 
For instance, the SG derived from the caption (in~\autoref{fig:task}) can be represented: \{(\textit{car}), (\textit{street, winding}), (\textit{car, two}), (\textit{car, gray}), (\textit{building, small}), (\textit{building, on either side of, street})\}.}

We employ a language model~\cite{Li:2023:FACTUAL} to conduct SGP on both the VL model output and the given GT. First, we extend the vocabulary in~\cite{Anderson:2016:spice} related to subcategories of attributes, focusing on color and size, to ensure more detailed representation and analysis.
By analyzing the differences between the resulting SGs, this approach effectively identifies the specific tasks being assessed in each instance with respect to the VL model's ability to perform them (relevant or not), as well as evaluates its performance on those tasks. To compare candidate and reference SGs, the reference SG is first enhanced by incorporating a synonym set for more accurate matching~\cite{Li:2023:FACTUAL}. A binary matching operation $\otimes$ is then applied to finding True Positives (TP, matching tuples) between the two SGs, along with the False Negatives (FN, tuples present only in reference), and False Positives (FP, tuples present only in candidate). 
This process enables a text-level analysis of the alignment and discrepancies across various semantic categories (i.e., tasks), $t \in \mathcal{T} = \{\texttt{object}, \texttt{attribute}, \texttt{relation}, \texttt{color}, \texttt{count}, \texttt{size}\}$.

\vspace{1.1mm}\noindent\textit{Image-level Judgment}:
While the text-level comparison provides a basic evaluation of the candidate and reference alignment for each task, the inherent incompleteness of GTs can cause correct predictions to be misjudged due to overlooked elements in the original image. For instance, a GT of image in~\autoref{fig:task} was described as: \textit{A building-lined street with three lanes and light traffic,} omits a \textit{car}, leading to it being misclassified as an FP despite correct identification by the candidate.
To address this issue, we introduce an image-level correction using the external steering model CLIP~\cite{Radford:2021:clip}. 
For each FP element, a descriptive sentence is generated using predefined templates. For instance, for an object FP like \textit{car}, the sentence would be: \textit{a photo of a car}. The CLIP image and text encoder are subsequently utilized to generate embeddings for both the original image and the corresponding generated sentence. The similarity between the embeddings is computed, and elements from the original FP set (i.e., $FP_{raw}$) that exceed a similarity threshold of 0.25, are identified. The threshold is set to balance modest single-element similarity with the image, preventing false negatives while ensuring a proper selection standard. These elements are removed from $FP_{raw}$ and assigned to the correct set, defined as $AC = FP_{raw} - FP_{new}$, where $FP_{new}$ is an updated tuple that reduces misclassifications.
The updated results provide a more precise view of the VL model's performance across tasks, offering insights into its ability to handle diverse tasks. Additionally, they allow for the computation of the task-based distribution from multiple dimensions, enabling a more comprehensive understanding of the model's behavior.

\vspace{1.2mm}\noindent\textbf{Data Transformation and Visual Encoding}: 
Building upon the refined results obtained from the image-level judgment, the process of visualization in~\autoref{fig:teaser}~(B) consists of two parts: data transformation and corresponding visual encoding. First, the data in the results of each task is transformed into several values that capture various aspects of the VL model's performance to interpret model behavior. 
For a dataset containing $N$ instances, is represented by $N\times M$ tuples in the format $(I_i, Cap_i,GT_{i,j}, t)$ where $I_i$ denotes an image, $Cap_i$ is the model output, $GT_{i,j}$ is the $j$-th GT ($j \in M$). 
\revise{For each instance, we assign a label $Atm_{i,t}$ to indicate whether the model attempts a specific task. If $|TP_{i,j,t}| + |FP^{raw}{i,j,t}| \neq 0$, then $Atm{i,t}=1$, meaning the task is attempted; otherwise, $Atm_{i,t}=0$. The error rate and attention-shifting rate are computed only when $Atm_{i,t}=1$. Besides, $Cnt_{t} = \sum^N_{i=1} \mathbb{I}(Atm_{i,t} = 1)$ represents the total samples where the model attempts the task.}

The error rate, $Err_{i,j,t}$, quantifies the proportion of erroneous elements produced by the model within a specific task:
\revise{\begin{equation}
\resizebox{.38\textwidth}{!}{$
Err_{i,j,t}=\frac{|FP^{new}_{i,j,t}|}{|TP_{i,j,t}|+|FP^{raw}_{i,j,t}|}; \quad Err_{i,t}=\frac{\sum^M_{j=1}Err_{i,j,t} \cdot \mathbb{I}(Atm{i,t}=1)}{\sum^M_{j=1}\mathbb{I}(Atm{i,t}=1)}.
$}
\end{equation}}
The attention-shifting rate, $Sf_{i,j,t}$, quantifies the model's attention shifts in focus from elements unique to $GT_{i,j}$ (i.e., $FN_{i,j,t}$) to elements uniquely and correctly identified within $Cap_i$ (i.e., $AC_{i,t}$) for image $I_i$:
\revise{\begin{equation}
\resizebox{.41\textwidth}{!}{$
     Sf_{i,j,t}=
\begin{cases}
\frac{|AC_{i,j,t}|}{|AC_{i,j,t}|+|FN_{i,j,t}|}, &\text{if denom.}\neq0\\
0,&\text{ otherwise} 
\end{cases};
\quad Sf_{i,t}=\frac{\sum^M_{j=1}Sf_{i,j,t} \cdot \mathbb{I}(Atm{i,t}=1)}{\sum^M_{j=1}\mathbb{I}(Atm{i,t}=1)}.
$}
\end{equation}}
Additionally, the sensitivity, $Sen_{i,j,t}$, represents the model's tendency to engage in a specific task among these tasks: 
\begin{equation}
\resizebox{.10\textwidth}{!}{$
Sen_{i,t}=\frac{|SG_{Cap_{i,t}}|}{|SG_{Cap_i}|},
$}
\end{equation}
where $SG_{Cap_i,t}=FP^{raw}_{i,j,t} \cup TP_{i,j,t}$. 

At the dataset level, the metrics $Err_t$, $Sf_t$, and $Sen_t$ for each task as following formulations:
\begin{equation}
\resizebox{.45\textwidth}{!}{$
Err_t=\frac{\sum^N_{i=1}Err_{i,t}\cdot\mathbb{I}(Atm_{i,t}=1)}{Cnt_t}; \quad Sf_t=\frac{\sum^N_{i=1}Sf_{i,t}\cdot\mathbb{I}(Atm_{i,t}=1)}{Cnt_t}; \revise{\quad Sen_t=\frac{\sum^N_{i=1}Sen_{i,t}\cdot\mathbb{I}(Atm_{i,t}=1)}{N}.}
$}
\end{equation}
The above metrics offer a comprehensive evaluation from multiple levels and aspects for each task, where $\mathbb{I}(\cdot)$ is the indicator function.

Based on the defined metrics, the task-driven inspection view organizes the task-based summary panel in a list-based layout, displaying $Cnt_t$, $Err_t$, $Sf_t$, and $Sen_t$ as colored bars. For each metric within the task, two aligned bars represent the selected corruption-severity level and the original input, enabling direct comparative analysis. \revise{Besides, density plots of $Err_t$, $Sf_t$, and $Sen_t$ of corrupted inputs are included to analyze their distributions, providing deeper insights into the impact of corruption on the VL model's behavior from the task perspective.}

\subsection{In-depth Visual Exploration of Data Patterns}\label{view:pattern}
At the most detailed level of analysis, this component focuses on uncovering salient data patterns for the selected task and analyzing instances within those patterns. When task $t$ is selected for further exploration, the method identifies patterns among instances where $Atm_{i,t} = 1$, based on two primary constraints: images and captions with higher visual and textual similarity are positioned closer together, and captions with lower similarity to their corresponding GT are similarly grouped. These constraints enable a summarization of model behavior based on these data patterns, highlighting \textbf{semantically similar} instances that consistently exhibit \textbf{poor performance}. Moreover, the identified patterns and detailed instance-level insights are employed as data augmentation samples to improve robustness, addressing \textbf{R4} in alignment with \textbf{DC3}.

\begin{figure}[!t]
  \centering
  \includegraphics[width=\linewidth]{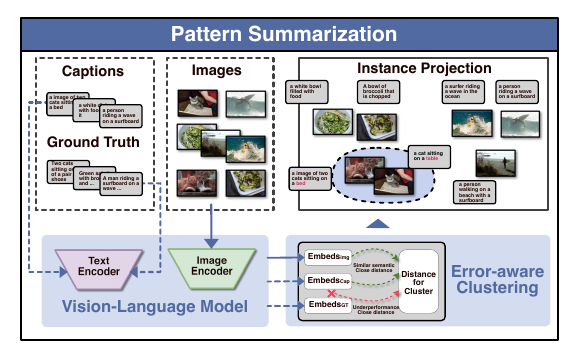}
  \caption{The approach of pattern discovery through error-aware clustering.}
  \label{fig:slice}
 \vspace{-6.2mm}
\end{figure}

\vspace{1.2mm}\noindent\textbf{Data Pattern Discovery and Analysis.}
To quantify the similarity between instances and uncover data patterns for further analysis and augmentation, a distance calculation method is introduced, guided by the aforementioned constraints base on~\cite{Eyuboglu:2022:domino}, shown in~\autoref{fig:slice}.
For two instances $i$ and $k$, their distance (i.e., $dist(i,k)$) is computed as follows:
First, the selected VL model image encoder (i.e., $f_{img}(\cdot)\in\mathbb{R}^{patch \times d_{img}}$) and text encoder (i.e., $f_{txt}(\cdot)\in\mathbb{R}^{\text{max\_length} \times d_{txt}}$) are applied to encode the image, caption, and GT, respectively.
To the first constraint, to compute the distances between images and between captions: Given the distance of Images (i.e., $dist_{I}(i,k)$) of $I_i$ and $I_k$, as well as the distance of Captions (i.e., $dist_{Cap}(i,k)$)  of $Cap_i$ and $Cap_k$:
\begin{equation}
\resizebox{.45\textwidth}{!}{$
dist_{img}(i,k) = 1-\texttt{cos}(E^{img}_i, E^{img}_k); \quad dist_{Cap}(i,k) = 1-\texttt{cos}(E^{Cap}_i, E^{Cap}_k),
$}
\end{equation}
where $E^{img}\in\mathbb{R}^{d_{img}}$, $E^{Cap}\in\mathbb{R}^{d_{txt}}$ and $E^{GT_j}\in\mathbb{R}^{d_{txt}}$ represents the maximum embedding (i.e., $max(\cdot)$) extracted from the $f_{img}(\cdot)$ and $f_{txt}(\cdot)$. 
\revise{To the second constraint, compute $\Delta$ as the difference between each caption and its GTs. The quality distance $dist_{qual}(i,k)$ ensures instances with similar error patterns are closer, measuring whether the ``direction of deviation'' between two caption-GT pairs is similar:
\begin{equation}
\resizebox{.34\textwidth}{!}{$
\Delta_i = E^{Cap}_i-\frac{\sum^{M}_{j=1}E^{GT}_{i,j}}{M}; \quad dist_{qual}(i,k)=1-\texttt{cos}(\Delta_i,\Delta_k),
$}
\end{equation}
where $\texttt{cos}(\cdot,\cdot)$ is the cosine similarity.}
Based on these two constraints, the distance between two instances can be computed as:
\begin{equation}
\resizebox{.42\textwidth}{!}{$
dist(i,k) = \frac{dist_{img}(i,k)+dist_{Cap}(i,k)}{2} \times (1-\alpha) + dist_{qual}(i,k) \times \alpha,
$}
\end{equation}
where $\alpha\in[0,1]$  is a hyperparameter balancing raw similarity and alignment quality. 
\revise{To align with the purpose discussed in~\autoref{discuss}, we set $\alpha=0.1$, and further validated it through consistency and cluster quality analysis in the supplementary material. 
}

\vspace{1.1mm}\noindent\textit{Layout and Visual Encoding}: 
To discern the underlying patterns among instances with similar behaviors, we utilize clustering to delineate their groupings based on the defined distance $dist(i,k)$. Specifically, we employ the HDBSCAN algorithm~\cite{McInnes:2017:hdbscan}, which operates on pairwise generation similarities, to identify cohesive clusters of instances. This approach is critical for capturing both local and global relationships within the data, offering a structured perspective on how instances contribute to similar outcomes in the VL model while effectively filtering out outliers, which either lack semantic similarity or exhibit underperformance that is not sufficiently representative, to facilitate valuable pattern discovery and analysis.
Projection techniques are applied to manage the complexity of instances combining textual and visual information, which leads to high-dimensional embeddings. Using UMAP~\cite{McInnes:2018:UUM}, these embeddings are projected into a two-dimensional space, enabling clearer visualization of relationships and patterns in a scatterplot.
Clustering results are further encoded with color for each dot using a categorical color scheme, emphasizing distinctions between patterns and highlighting internal relationships within clusters, offering a more structured and intuitive understanding of the data.
Besides, to mitigate and enhance labeled clusters with more visual guidance, an additional design is to draw a contour map for each labeled cluster using the kernel density estimation method, shown in~\autoref{fig:teaser}~(C1). This is intended to reveal the density of the dots, and the density levels are represented by the lightness of the filled contour colors and the stroke width of the contour lines.
To provide a more meaningful semantic description of each cluster's patterns, the most frequent element in the captions within the selected task is displayed at the cluster's centroid.

\vspace{1.1mm}\noindent\textit{Interaction}:
When hovering over a dot, a tooltip is activated to display basic instance information (i.e., image ID, label, error rate, shifting rate, and sensitivity). Similarly, hovering over captions on a centroid reveals aggregated information at the cluster level.
The interface supports lasso selection in the projection scatterplot~(\autoref{fig:teaser}~(C1)), for selecting instances for detailed inspection and analysis.

\begin{figure}[!t]
  \centering
  \includegraphics[width=0.99\linewidth]{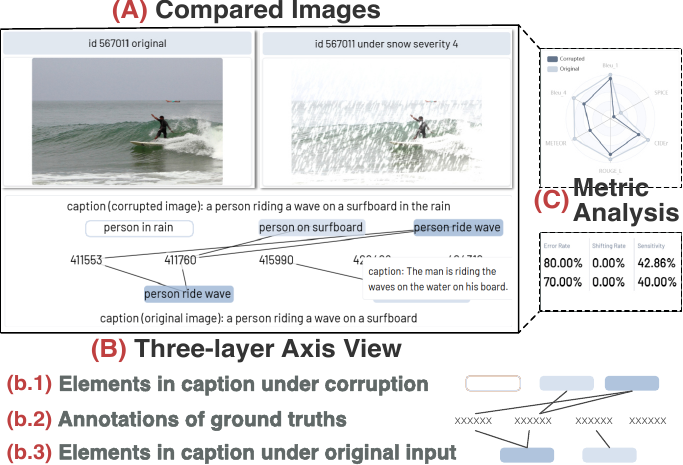}
  \caption{The instance inspection view is structured by A) compared images part; B) three-layer axis view part, and C) metric analysis part.}
  \label{fig:slice:glyph}
 \vspace{-6.8mm}
\end{figure}

\begin{figure*}[!htbp]
\setlength{\abovecaptionskip}{0.17mm} 
    \centering
    \includegraphics[width=0.95\textwidth]{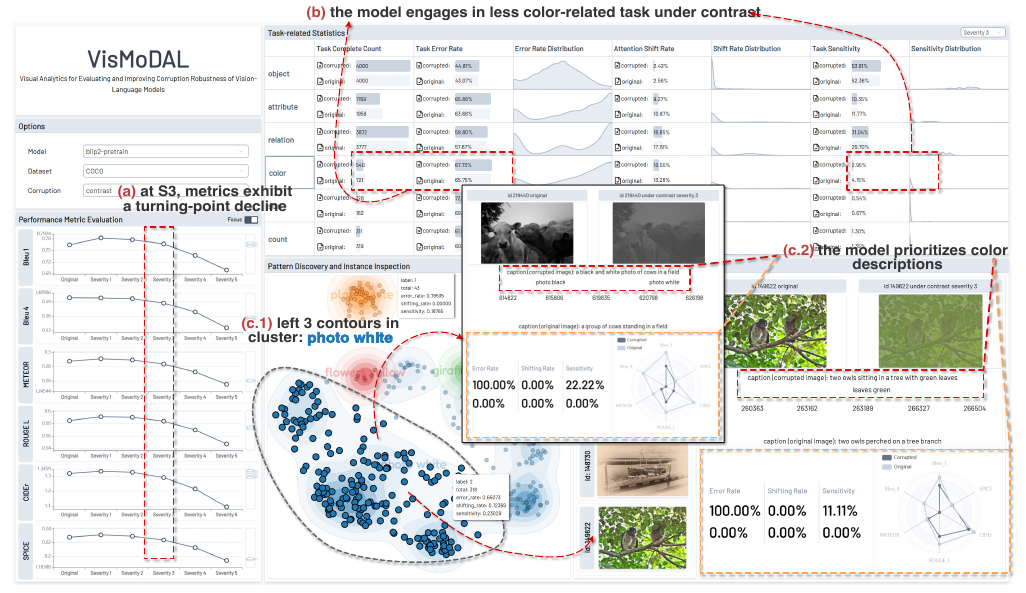}
    \caption{This case explores a semantically less relevant corruption: contrast, guiding analysts through: a) identifying performance shifts, b) observing reduced but concentrated sensitivity of color, and c.1–c.2) analyzing patterns from key clusters, suggesting increased sensitivity for specific instances.}
        \label{fig:case2}
 \vspace{-5mm}
\end{figure*}

\vspace{1.2mm}\noindent\textbf{Detailed Instance Inspection and Analysis.}
Based on the selected samples with similar patterns, analysts can examine the behavior of individual instances and compare their performance within the selected task using GTs. The instance-level view is organized into two interconnected panels. The right panel displays the images corresponding to the selected instances, while the left panel, upon selecting an image, reveals more detailed visual attributes of the corresponding instance. This design provides fine-grained insights to evaluate the VL model's behavior when processing the instance to complete the task. Additionally, it serves as guidance for identifying these patterned instances as an effective subset for data augmentation in training.

\vspace{1.1mm}\noindent\textit{Visualization and Interaction}: 
To support the comparison and validation of matched elements for a selected instance, the first part of the left panel presents the original and corrupted images. The second part employs a three-layer axis layout~(\autoref{fig:slice:glyph}): Task-relevant elements extracted from captions under the specified corruption severity level are displayed as individual, color-coded tags on the top layer of the axis view. These colors represent the frequency of occurrence of each element in the GT.
If a tag corresponds to an element present in the GT, it is connected to the respective GT card with a line on the middle layer of the axis view, enabling straightforward comparison.
The bottom layer of the axis view represents elements from the caption associated with the original input. Similarly, each card in this layer is color-coded, and connections are drawn to the GT cards based on the same linking logic as in the top layer.
This hierarchical and interactive design enables an intuitive and detailed comparison of the relationships among captions generated by the VL model under corruption, captions from the original condition, and the GTs. Additionally, hovering over any GT tag reveals detailed information about its content, allowing analysts to examine the alignment between the VL model output and the GT.
$Err_{i,t}$, $Sf_{i,t}$, and $Sen_{i,t}$ are displayed in the third section, along with six standard performance metrics represented using a radar chart. Each axis of the radar chart corresponds to a specific metric, with the results for the selected corruption type and the original inputs overlaid and differentiated by distinct colors for clarity.

\section{Evaluation}
To demonstrate the framework's effectiveness, two case studies were conducted with the involvement of domain experts, and their feedback was collected through post-interviews to support further discussion. Subsequently, quantitative evaluations were performed to provide additional validation of the framework's utility in enhancing robustness.
The VisMoDAl framework is implemented using a browser-server architecture, leveraging PyTorch for model processing. Python Flask for server-side logic, and Vue3 and D3.js for client-side development. For the case studies and quantitative analysis, the widely-recognized COCO dataset~\cite{Lin:2014:MCC} is employed, which is partitioned into two subsets to serve as a benchmark for case studies and quantitative evaluation.

\vspace{1.2mm}\noindent\textbf{Participants.} 
\revise{The case studies were conducted by two experts specializing in multi-modal robust learning, including E2 (a co-author in this work) and an additional expert, E3, recruited via the university mailing list.} E3, a senior Ph.D. student in robust deep learning, focuses on research areas including transferable adversarial attacks and backdoor attacks on multi-modal models.

\vspace{1.2mm}\noindent\textbf{Procedure.}
During the interview, we first introduce the background and analytical tasks, followed by presenting the interface and visualization design of the framework. This process allowed the experts to become familiar with the framework interface and gain the foundational knowledge necessary for the system setup.
\revise{In the next step, as the participant first engaged with our work, E3 utilized VisMoDAl to conduct a comprehensive workflow analysis on a common and representative corruption type. This process involved evaluating model robustness and identifying specific dataset sub-samples, aiming to develop strategies for enhancing robustness under typical real-world corruption scenarios. Subsequently, E2 conducted an analysis focused on a less semantic corruption type, which provided fresh insights into VL model behavior through a task-driven approach, uncovering detailed patterns that spanned from dataset-level evaluation to instance-level understanding, thereby validating the effects of this less conventional corruption type. After the completion of both cases, the processes and findings from each expert were shared with the other to ensure counterbalance and foster mutual understanding.
Separate open discussion sessions were arranged after the case analyses, allowing the experts to freely explore the framework and provide constructive feedback. Their comments regarding the framework's design, functionality, and usability were collected. Each interview lasted between 1 and 1.5 hours.}

\subsection{Case 1: Robustness Analysis for Improvement}
\revise{We invited E3 to use VisMoDAl with the entire analysis workflow.} The first case study focuses on task-driven behavior understanding to enable analysis and pattern exploration, aiming to identify and select samples for augmentation to enhance model robustness,~\autoref{fig:teaser}.

\vspace{1.2mm}\noindent\textbf{Overall Performance Metrics Evaluation (DC1).}
E3 began by selecting the snow corruption type, which is a common and particularly challenging issue for VL models in real-world scenarios. E3 then examined the overall performance metrics displayed as line charts. As the corruption severity increased from level 3 to 5, a significant performance degradation at severity level 4 can be observed. Specifically, BLEU-1 dropped to 0.7089, BLEU-4 to 0.3177, and ROUGE\_L to 0.5194, indicating worse performance across these metrics at severity level 4 (S4). Interestingly, a partial recovery was observed at severity level 5,~\autoref{fig:teaser}~(A2). This trend drew E3's attention to S4 as the critical point to understand the underlying reasons for the performance drop under snow corruption.

\vspace{1.2mm}\noindent\textbf{Task-driven Model Behavior Understanding (DC2).}
E3 then analyzed \autoref{fig:teaser}~(B), comparing task-specific measurement bars between snow corruption at S4 and the original input, observing an increase in relation-related instances, rising from 3,777 to 3,938, alongside a notable rise in the error rate from 57.87\% to 72.44\%. 
It prompted E3 to suggest that snow corruption may considerably degrade the model's performance on relation-related tasks while also revealing an increased tendency of the VL model to output relation-related content,~\autoref{fig:teaser}~(B). E3 argued that snow corruption at S4 affected the sensitivity of relation-related tasks, and results showed an increase in sensitivity under this condition, \revise{with the value increasing from 29.70\% to 33.08\%, \autoref{fig:teaser}~(B).} Comparing sensitivity distributions across tasks, E3 observed that relation-related tasks had higher values in the upper range, suggesting that snow corruption at S4 prompts greater engagement in these tasks but reduces accuracy.
Besides, the attention-shifting rate dropped from 17.19\% to 13.14\%, reflecting fewer correct attention shifts. This decline aligns with higher error rates in relation tasks and overall VL model degradation under snow corruption.
\revise{Compared to other tasks, relation showed significant change, leading E3 to hypothesize that snow primarily impacts model ability on relation-related tasks.}

\vspace{1.2mm}\noindent\textbf{Data Pattern Exploration and Instance-level Analysis (DC3).}
\revise{Building on previous findings, E3 clicked on the relation to analyze instances, revealing six clusters in the projection view~(\autoref{fig:teaser}~(C1)). Examining the clusters, E3 found that the red and orange ones, linked to skiing and surfing respectively, showed the highest sensitivity and relatively high error rates, both associated with snow-related images. These samples indicated that the model’s sensitivity to snow-related relations stemmed from data characteristics~(snow and rain in \autoref{fig:teaser}~(c.1.1) and \autoref{fig:slice:glyph}) rather than corruption. E3 then focused on the primary green cluster, the largest and exhibiting a higher error rate, labeled "\textit{snow fall from ceiling}," and try to identify the pattern within this subset had a more significant impact on model performance~(\autoref{fig:teaser}~(c.1.2)).}
E3 selected green dots and found these samples primarily concentrated in two contours (i.e., categories), as guided by contours: outdoor transportation and residential indoor scenes. E3 then examined these images. For image \#14038, snow disrupted the model's perception of indoor object relationships, interpreting the scene as an empty house in~\autoref{fig:teaser}~(c.2.1). In \#23666, snow shifted relational focus to emphasize connections with snow, misclassifying a bathtub as a sink, shown as~\autoref{fig:teaser}~(c.2.2). For outdoor scenes~(\autoref{fig:teaser}~(c.2.3-2.4)), in Image \#6723, the model's attention diminished the relational context between the road and the building. Similarly, in Image \#168330, the model's focus shifted from the clock tower at the street corner to a more high-level description, such as being in front of a building. \revise{These results highlight how snow significantly affects the model's ability to maintain accurate relational reasoning, leading to attention shifts and potential errors beyond snow-related relations.}
By examining the instances from these three clusters, E3 identified different behaviors in relation-related tasks and errors, and assessed their impact on the model's performance in such tasks. Considering the representativeness of the green cluster, E3 used 132 images from this cluster to enhance robustness against snow via data augmentation and validate the hypothesis in~\autoref{sec:quan}.

\begin{figure}[!htbp]
\setlength{\abovecaptionskip}{0.20cm} 
    \centering
    \includegraphics[width=0.49\textwidth]{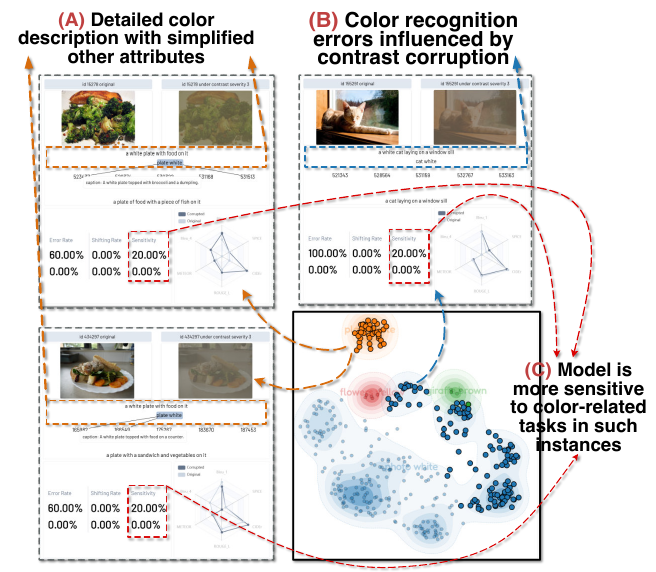}
    \caption{Based on the hypothesis of the model's color sensitivity under contrast corruption, analysts examined two other patterns: A) detailed color descriptions with simplified attributes, B) color recognition errors, validating C) effects from rich, high-contrast inputs.}
        \label{fig:case2.2}
     \vspace{-6.5mm}
\end{figure}

\subsection{Case 2: Robustness Evaluation with Validation}
Case 2 evaluates VL model performance through task-driven analysis, focusing on understanding the impact of a semantically less relevant corruption, unlike snow. \revise{The case was conducted with E2.}

\vspace{1.2mm}\noindent\textbf{Overall Performance Metrics Evaluation (DC1).}
Contrast refers to a common non-semantic corruption caused by transmission or medium issues, resulting in excessive contrast. E2 in this case, observed that the metrics for S0-S2 remained relatively stable, while a decline was evident at S3 across all six metrics displayed in the line charts. Notably, the scores for METEOR, CIDEr, and SPICE dropped from 0.2893, 1.3249, and 0.2293 at S2 to 0.2837, 1.2900, and 0.2238 at S3~(\autoref{fig:case2}~(a)), respectively, leading to the selection of S3 for further investigation.

\vspace{1.2mm}\noindent\textbf{Task-driven Model Behavior Understanding (DC2).}
E2 analyzed the task-related measurement view,~\autoref{fig:case2}~(b), to identify how the model's performance on specific tasks was affected by contrast. \revise{When comparing the results to those from the original input, it was observed that the contrast at S3 led to a reduction in the number of instances where the VL model-generated outputs involving color-related content decreased to 540, and the model's sensitivity to color-related tasks dropped to 2.95\%.} Additionally, the sensitivity distribution was concentrated in the lower range. 
\revise{The reduction in color sensitivity despite its relationship to contrast was unexpected. Thus, E2 hypothesized that only images with specific prominent features become more sensitive to color under contrast conditions, making them more vulnerable to corruption and contributing to the increased error rate.}
To further understand this phenomenon, E2 focused on analyzing color-related patterns and individual cases to uncover potential underlying causes.

\vspace{1.2mm}\noindent\textbf{Data Pattern Exploration and Instance-level Analysis (DC3).}
\revise{Based on the projection view, E2 first investigated the largest cluster with high error rate}, "\textit{photo white}," which included several contours. In the left contours, instances such as the black-and-white Image \#219440 demonstrated that contrast corruption shifted the model's behavior from ignoring color to emphasizing it~(\autoref{fig:case2}~(c.2)). Similarly, for Image \#149622, the model displayed heightened attention to green elements~(\autoref{fig:case2}~(c.2)). In contrast, the "\textit{leaves green}" element showed no GT connections and had high error rates, indicating that the GT annotations might lack sufficient granularity.
In other contours of the blue cluster, instances like Image \#155291 revealed that high contrast led to color misclassification, with the brown fur of the cat being perceived as white~(\autoref{fig:case2.2}~(B)). This observation suggested that contrast generally reduced sensitivity, except in cases of vibrant or highly contrasted images, where the model demonstrated heightened sensitivity compared to the original condition.
\revise{Within the orange "\textit{plate white}" cluster, which had the highest error rate}, food-related images like \#434297 and \#15278 exhibited a recurring trend (\autoref{fig:case2.2}~(A)): contrast corruption caused the model to shift from producing detailed food-related outputs to focusing on secondary features, such as the plate's color. E2 concluded and validated that contrast corruption amplifies existing high contrast in the data, significantly influencing the model's focus and predictions.

\subsection{Expert Feedback}
We summarize experts’ feedback from the perspective of the framework, visualization, and limitations.

\vspace{1.2mm}\noindent\textbf{Framework.}
Experts recognized the framework's effectiveness and practicality during the case studies. E3 highlighted its integration of data-driven evaluation, task-based behavior understanding, and visual analysis, which together enable a systematic identification of model vulnerabilities under various corruptions. E2 further emphasized the multi-dimensional structure of the framework, combining six standard metrics, six task categories, and three task-specific measurements to support comprehensive comparative analyses. 
Besides, E2 acknowledged the framework's multi-granular analysis capability, using task-related metrics and clustering for pattern discovery, which enhances interpretability, identifies data bias and GT insufficiencies, and provides insights to improve VL model robustness.

\vspace{1.2mm}\noindent\textbf{Visualization.}
The interactive visualization design received positive feedback for its efficiency in presenting diverse metrics, supporting multi-facet comparisons, and uncovering relationships among the VL model, instances, and tasks. E2 commended the intuitive task-driven visualization, which combines high-level statistical views with fine-grained instance inspection, supporting the exploration of model behavior. E2 further highlighted the framework's capability for comparative analysis at both overall and detailed levels through juxtaposed and superposed visualizations, aiding in the identification of variations between corrupted and original data. E3 emphasized the utility of clustering and projection methods in uncovering error patterns with similar behaviors, as well as their contribution to pattern discovery and informing data augmentation strategies for robustness enhancement.

\vspace{1.2mm}\noindent\textbf{Limitation.}
E3 observed the absence of integration with the training process but acknowledged that the extended duration of training renders it less practical for interactive systems. E2 identified limitations in using COCO, noting that its concise captions might constrain analysis accuracy, and recommended incorporating datasets with more detailed captions and additional models for broader comparisons. While currently applied to a single dataset and the foundational multi-modal BLIP-2, the framework is designed to support extension to any VL dataset and model, ensuring scalability.

\subsection{Quantitative Evaluation}\label{sec:quan}
\revise{Building on E3's analysis in Case 1, we collected samples for effective data augmentation. To validate their utility and further examine the performance of VisMoDAl, we evaluated the robustness of various models, including the pretrained BLIP-2 and four models specifically augmented with Low-Rank Adaptation (LoRA) to address snow corruption, while keeping all other configurations consistent across models. 
The evaluation included: the pretrained BLIP-2 (the model can be selected in our framework for analysis),
the BLIP-2 fine-tuned on the COCO training dataset with snow corruption (snow S4), 
and BLIP-2 fine-tuned on 132 samples collected from the first case study with snow corruption (VisMoDAl), alongside two control groups:
132 randomly selected samples (random) and 132 samples with the highest error rate (error\_rate), both from the projection view.
The robustness of these models was validated on the COCO test dataset under snow at S4 (1,000 images).
}  
All models trained with DA show performance improvements across all metrics for the specific corruption~(\autoref{table1}). Notably, the model trained on samples identified by VisMoDAl in Case 1 outperforms others in BLEU1, BLEU4, ROUGE\_L, and CIDEr. Although METEOR and SPICE are slightly lower than full-dataset augmentation, the training set size of 132 highlights the strategy's efficiency, demonstrating that VisMoDAl effectively identifies sub-datasets for targeted augmentation, enhancing robustness against specific corruption.
\revise{The evaluations for three other representative corruption types are included in the supplemental material.}

\begin{table}[!t]
\caption{The results of quantitative evaluation}
\label{table1}
\renewcommand\arraystretch{1.8}
\resizebox{0.49\textwidth}{!}{
\begin{tabular}{cccccccc}
\toprule
Model           & BLEU1          & BLEU4           & METEOR         & ROUGE\_L        & CIDEr          & SPICE & Data Size \\
\midrule
BLIP-2          & 0.6932         & 0.3034          & 0.2449          & 0.5089          & 1.0419          & 0.1788 & -            \\  
BLIP-2 (snow S4) & 0.7070          & 0.3113          & \textbf{0.2487}          & 0.5137          & 1.0797          & \textbf{0.1805} & 4000        \\
BLIP-2 (random) & 0.7072  &  0.3142  &  0.2481 & 0.5144  & 1.0779  & 0.1794  & \textbf{132} \\
BLIP-2 (error\_rate)    &  0.7087  & 0.3163  & 0.2484  & 0.5165  & 1.0754  & 0.1796   & \textbf{132} \\
BLIP-2 (\textbf{ours})   & \textbf{0.7098} & \textbf{0.3173} &  \underline{0.2485} & \textbf{0.5187} & \textbf{1.0813} & \underline{0.1801} & \textbf{132} \\
\bottomrule
\end{tabular}
}
\vspace{-6.8mm}
\end{table}

\section{Discussions and Conclusion}
\label{discuss}

In this paper, we introduce VisMoDAl, a visual analytics framework designed to evaluate and enhance the robustness of VL models under corruption in a task-driven manner. The framework adopts a multi-level design, enabling analysts to investigate model behavior from various perspectives, including performance metrics, task-specific measures, data pattern exploration, and instance-level analysis. It moves beyond traditional numerical result-based evaluations, fostering a deeper semantic understanding of model behavior by establishing connections to task-related contexts. The utility and effectiveness of the proposed framework are demonstrated through case studies and expert feedback.

\vspace{1.2mm}\noindent\textbf{Scalability.} We discuss the scalability of our framework from two aspects: data preprocessing and visual design.

\vspace{1.1mm}\noindent\textit{Data Preprocessing}:
The framework's scalability is notably limited by the computational demands of preprocessing and data transformation. Preprocessing the COCO-val17 dataset (5,000 images) to generate 80 corrupted variants~\cite{Hendrycks:2018:BNN, Qiu:2024:BRO} takes approximately 12 hours. 
\revise{The subsequent data transformation phase, including SGP, image-level judgment, and error-aware distance calculation, adds 3 hours to construct 81 datasets.} 
This emphasizes the trade-offs between preprocessing and data transformation in scaling the framework.

\vspace{1.1mm}\noindent\textit{Visual Design}:
Our framework incorporates scatterplot-based designs, with scalability challenges partially addressed through clustering based on error-aware distances. This method filters out non-essential samples that do not align with the primary evaluation objectives or significantly impact model underperformance,  thereby reducing dataset size and computational complexity. Additionally, the multi-granular visual analytics framework adheres to the ``overview+detail'' principle, leveraging multiple tasks and summarized patterns enriched with semantics. This design enables analysts to efficiently evaluate and derive meaningful insights into model performance across large-scale datasets.

\vspace{1.2mm}\noindent\textbf{Generalizability.}
We demonstrate the functionality of our system using BLIP-2, a foundational and baseline VL model, evaluated on the widely recognized COCO dataset. However, it is important to note that the system is not restricted to the COCO and can be extended to evaluate the corruption robustness of VL models with a variety of image-text datasets. 
For instance, FineCapEval~\cite{Cho:2022:FGI} is a dataset designed to comprehensively evaluate image captions, addressing the issue of insufficiently descriptive GTs in existing datasets.
Besides, the framework generalizes to more corruption types if definitions and processing follow the protocol.

\vspace{1.2mm}\noindent\textbf{Hyperparameter Setting.}
The threshold for image-level judgment was set to 0.25 after testing values within 0.1–0.5. Lower settings caused nearly all inputs to be classified as similar to the image, while higher settings led to single caption elements, even when reconstructed into sentences, being inaccurately judged as unrelated. 
\revise{Similarly, the hyperparameter $\alpha$ for distance calculation need to align with discovery principles~\cite{Eyuboglu:2022:domino}. Higher values grouped poorly performing instances, whereas lower values, emphasizing semantic similarity, were prioritized. To balance this with coherence, $\alpha = 0.1$ was selected. This choice was further validated through consistency and cluster quality analyses under hyperparameter variation, as detailed in the supplemental material.}

\vspace{1.2mm}\noindent\textbf{Limitations and Future Work.}
The current demonstration includes various corruption types, but enumerating all possible real-world scenarios is inherently infeasible, which highlights the need for further discussion and extension in our framework. \revise{The scale of experts recruited for the case study needs to be expanded.} Additionally, the task categories presented are not comprehensive enough to examine all dimensions of VL model performance. We plan to explore additional task categories beyond those listed in Section~\ref{sec:task-driven-vl-model-perf-evaluation}. Given the diversity of VL models across different application scenarios, a promising direction for testing the applicability of our framework is to apply it to other tasks, such as visual grounding and visual question answering. \revise{The current evaluation is limited to COCO and a foundation model, BLIP-2. Future work will focus on expanding the dataset (e.g., FineCapEval~\cite{Cho:2022:FGI}) to cover broader downstream tasks and incorporating more state-of-the-art large VL models to enhance the framework's applicability.}


\section*{Supplemental Materials}
\label{sec:supplemental_materials}

The supplemental materials include 1) a demo video of our framework and the corresponding subtitle file of the video; 2) a PDF providing an analysis of hyperparameters and quantitative evaluations across three other representative corruption types.

\acknowledgments{%
This work was supported in part by the National Natural Science Foundation of China (No. 62202217), Guangdong Basic and Applied Basic Research Foundation (No. 2023A1515012889), and Guangdong Key Program (No. 2021QN02X794).%
}

\bibliographystyle{abbrv-doi-hyperref}

\bibliography{reference}

\begin{thebibliography}{10}

\bibitem{Aflalo:2022:VIA}
E.~Aflalo, M.~Du, S.-Y. Tseng, Y.~Liu, C.~Wu, N.~Duan, and V.~Lal.
\newblock Vl-interpret: An interactive visualization tool for interpreting vision-language transformers.
\newblock In {\em Proc.\ CVPR}, pp. 21374--21383, 2022. \href{https://doi.org/10.1109/CVPR52688.2022.02072}
{doi: {{%
10\hspace{.1pt}\discretionary{.}{%
}{.}\hspace{.4pt}1109\discretionary{/}{%
}{/}CVPR52688\hspace{.1pt}\discretionary{.}{%
}{.}\hspace{.4pt}2022\hspace{.1pt}\discretionary{.}{%
}{.}\hspace{.4pt}02072}}}


\bibitem{Anderson:2016:spice}
P.~Anderson, B.~Fernando, M.~Johnson, and S.~Gould.
\newblock Spice: Semantic propositional image caption evaluation.
\newblock In {\em Proc.\ ECCV}, pp. 382--398, 2016. \href{https://doi.org/10.1007/978-3-319-46454-1_24}
{doi: {{%
10\hspace{.1pt}\discretionary{.}{%
}{.}\hspace{.4pt}1007\discretionary{/}{%
}{/}978\discretionary{%
}{-}{-}3\discretionary{%
}{-}{-}319\discretionary{%
}{-}{-}46454\discretionary{%
}{-}{-}1\_24}}}


\bibitem{Anderson:2018:task}
P.~Anderson, X.~He, C.~Buehler, D.~Teney, M.~Johnson, S.~Gould, and L.~Zhang.
\newblock Bottom-up and top-down attention for image captioning and visual question answering.
\newblock In {\em Proc.\ CVPR}, pp. 6077--6086, 2018. \href{https://doi.org/10.1109/CVPR.2018.00636}
{doi: {{%
10\hspace{.1pt}\discretionary{.}{%
}{.}\hspace{.4pt}1109\discretionary{/}{%
}{/}CVPR\hspace{.1pt}\discretionary{.}{%
}{.}\hspace{.4pt}2018\hspace{.1pt}\discretionary{.}{%
}{.}\hspace{.4pt}00636}}}


\bibitem{Baltruvsaitis:2019:MML}
T.~Baltrušaitis, C.~Ahuja, and L.-P. Morency.
\newblock Multimodal machine learning: A survey and taxonomy.
\newblock {\em IEEE Transactions on Pattern Analysis and Machine Intelligence}, 41(2):423--443, Feb. 2019. \href{https://doi.org/10.1109/TPAMI.2018.2798607}
{doi: {{%
10\hspace{.1pt}\discretionary{.}{%
}{.}\hspace{.4pt}1109\discretionary{/}{%
}{/}TPAMI\hspace{.1pt}\discretionary{.}{%
}{.}\hspace{.4pt}2018\hspace{.1pt}\discretionary{.}{%
}{.}\hspace{.4pt}2798607}}}


\bibitem{Banerjee:2005:meteor}
S.~Banerjee and A.~Lavie.
\newblock {METEOR}: An automatic metric for {MT} evaluation with improved correlation with human judgments.
\newblock In {\em Proc.\ ACL Workshop}, pp. 65--72, June 2005.

\bibitem{Bednarek:2020:ORO}
M.~Bednarek, P.~Kicki, and K.~Walas.
\newblock On robustness of multi-modal fusion—robotics perspective.
\newblock {\em Electronics}, 9(7):1152, 2020.

\bibitem{Bhojanapalli:2021:URO}
S.~Bhojanapalli, A.~Chakrabarti, D.~Glasner, D.~Li, T.~Unterthiner, and A.~Veit.
\newblock Understanding robustness of transformers for image classification.
\newblock In {\em Proc.\ ICCV}, pp. 10211--10221, 2021. \href{https://doi.org/10.1109/ICCV48922.2021.01007}
{doi: {{%
10\hspace{.1pt}\discretionary{.}{%
}{.}\hspace{.4pt}1109\discretionary{/}{%
}{/}ICCV48922\hspace{.1pt}\discretionary{.}{%
}{.}\hspace{.4pt}2021\hspace{.1pt}\discretionary{.}{%
}{.}\hspace{.4pt}01007}}}


\bibitem{Cao:2021:ATN}
K.~Cao, M.~Liu, H.~Su, J.~Wu, J.~Zhu, and S.~Liu.
\newblock Analyzing the noise robustness of deep neural networks.
\newblock {\em IEEE Transactions on Visualization and Computer Graphics}, 27(7):3289--3304, 2021. \href{https://doi.org/10.1109/TVCG.2020.2969185}
{doi: {{%
10\hspace{.1pt}\discretionary{.}{%
}{.}\hspace{.4pt}1109\discretionary{/}{%
}{/}TVCG\hspace{.1pt}\discretionary{.}{%
}{.}\hspace{.4pt}2020\hspace{.1pt}\discretionary{.}{%
}{.}\hspace{.4pt}2969185}}}


\bibitem{Chen:2024:CVeval}
C.~Chen, Y.~Guo, F.~Tian, S.~Liu, W.~Yang, Z.~Wang, J.~Wu, H.~Su, H.~Pfister, and S.~Liu.
\newblock A unified interactive model evaluation for classification, object detection, and instance segmentation in computer vision.
\newblock {\em IEEE Transactions on Visualization and Computer Graphics}, 30(1):76--86, 2024. \href{https://doi.org/10.1109/TVCG.2023.3326588}
{doi: {{%
10\hspace{.1pt}\discretionary{.}{%
}{.}\hspace{.4pt}1109\discretionary{/}{%
}{/}TVCG\hspace{.1pt}\discretionary{.}{%
}{.}\hspace{.4pt}2023\hspace{.1pt}\discretionary{.}{%
}{.}\hspace{.4pt}3326588}}}


\bibitem{Chen:2022:vis4annotation}
C.~Chen, J.~Wu, X.~Wang, S.~Xiang, S.-H. Zhang, Q.~Tang, and S.~Liu.
\newblock Towards better caption supervision for object detection.
\newblock {\em IEEE Transactions on Visualization and Computer Graphics}, 28(4):1941--1954, 2022. \href{https://doi.org/10.1109/TVCG.2021.3138933}
{doi: {{%
10\hspace{.1pt}\discretionary{.}{%
}{.}\hspace{.4pt}1109\discretionary{/}{%
}{/}TVCG\hspace{.1pt}\discretionary{.}{%
}{.}\hspace{.4pt}2021\hspace{.1pt}\discretionary{.}{%
}{.}\hspace{.4pt}3138933}}}


\bibitem{Chen:2021:oodanalyzer}
C.~Chen, J.~Yuan, Y.~Lu, Y.~Liu, H.~Su, S.~Yuan, and S.~Liu.
\newblock Oodanalyzer: Interactive analysis of out-of-distribution samples.
\newblock {\em IEEE Transactions on Visualization and Computer Graphics}, 27(7):3335--3349, 2021. \href{https://doi.org/10.1109/TVCG.2020.2973258}
{doi: {{%
10\hspace{.1pt}\discretionary{.}{%
}{.}\hspace{.4pt}1109\discretionary{/}{%
}{/}TVCG\hspace{.1pt}\discretionary{.}{%
}{.}\hspace{.4pt}2020\hspace{.1pt}\discretionary{.}{%
}{.}\hspace{.4pt}2973258}}}


\bibitem{Cheung:2024:augCV}
T.-H. Cheung and D.-Y. Yeung.
\newblock A survey of automated data augmentation for image classification: Learning to compose, mix, and generate.
\newblock {\em IEEE Transactions on Neural Networks and Learning Systems}, 35(10):13185--13205, 2024. \href{https://doi.org/10.1109/TNNLS.2023.3282258}
{doi: {{%
10\hspace{.1pt}\discretionary{.}{%
}{.}\hspace{.4pt}1109\discretionary{/}{%
}{/}TNNLS\hspace{.1pt}\discretionary{.}{%
}{.}\hspace{.4pt}2023\hspace{.1pt}\discretionary{.}{%
}{.}\hspace{.4pt}3282258}}}


\bibitem{Cho:2022:FGI}
J.~Cho, S.~Yoon, A.~Kale, F.~Dernoncourt, T.~Bui, and M.~Bansal.
\newblock Fine-grained image captioning with {CLIP} reward.
\newblock In {\em Proc.\ Findings of NAACL}, pp. 517--527, 2022. \href{https://doi.org/10.18653/v1/2022.findings-naacl.39}
{doi: {{%
10\hspace{.1pt}\discretionary{.}{%
}{.}\hspace{.4pt}18653\discretionary{/}{%
}{/}v1\discretionary{/}{%
}{/}2022\hspace{.1pt}\discretionary{.}{%
}{.}\hspace{.4pt}findings\discretionary{%
}{-}{-}naacl\hspace{.1pt}\discretionary{.}{%
}{.}\hspace{.4pt}39}}}


\bibitem{Cubuk:2019:autoaug}
E.~D. Cubuk, B.~Zoph, D.~Mané, V.~Vasudevan, and Q.~V. Le.
\newblock Autoaugment: Learning augmentation strategies from data.
\newblock In {\em Proc.\ CVPR}, pp. 113--123, 2019. \href{https://doi.org/10.1109/CVPR.2019.00020}
{doi: {{%
10\hspace{.1pt}\discretionary{.}{%
}{.}\hspace{.4pt}1109\discretionary{/}{%
}{/}CVPR\hspace{.1pt}\discretionary{.}{%
}{.}\hspace{.4pt}2019\hspace{.1pt}\discretionary{.}{%
}{.}\hspace{.4pt}00020}}}


\bibitem{Deng:2025:adversal}
D.~Deng, C.~Zhang, H.~Zheng, Y.~Pu, S.~Ji, and Y.~Wu.
\newblock Adversaflow: Visual red teaming for large language models with multi-level adversarial flow.
\newblock {\em IEEE Transactions on Visualization and Computer Graphics}, 31(1):492--502, 2025. \href{https://doi.org/10.1109/TVCG.2024.3456150}
{doi: {{%
10\hspace{.1pt}\discretionary{.}{%
}{.}\hspace{.4pt}1109\discretionary{/}{%
}{/}TVCG\hspace{.1pt}\discretionary{.}{%
}{.}\hspace{.4pt}2024\hspace{.1pt}\discretionary{.}{%
}{.}\hspace{.4pt}3456150}}}


\bibitem{Deng:2021:DLB}
Y.~Deng, T.~Zhang, G.~Lou, X.~Zheng, J.~Jin, and Q.-L. Han.
\newblock Deep learning-based autonomous driving systems: A survey of attacks and defenses.
\newblock {\em IEEE Transactions on Industrial Informatics}, 17(12):7897--7912, 2021. \href{https://doi.org/10.1109/TII.2021.3071405}
{doi: {{%
10\hspace{.1pt}\discretionary{.}{%
}{.}\hspace{.4pt}1109\discretionary{/}{%
}{/}TII\hspace{.1pt}\discretionary{.}{%
}{.}\hspace{.4pt}2021\hspace{.1pt}\discretionary{.}{%
}{.}\hspace{.4pt}3071405}}}


\bibitem{DeRose:2021:AFA}
J.~F. DeRose, J.~Wang, and M.~Berger.
\newblock Attention flows: Analyzing and comparing attention mechanisms in language models.
\newblock {\em IEEE Transactions on Visualization and Computer Graphics}, 27(2):1160--1170, 2021. \href{https://doi.org/10.1109/TVCG.2020.3028976}
{doi: {{%
10\hspace{.1pt}\discretionary{.}{%
}{.}\hspace{.4pt}1109\discretionary{/}{%
}{/}TVCG\hspace{.1pt}\discretionary{.}{%
}{.}\hspace{.4pt}2020\hspace{.1pt}\discretionary{.}{%
}{.}\hspace{.4pt}3028976}}}


\bibitem{Devlin:2019:BPT}
J.~Devlin, M.-W. Chang, K.~Lee, and K.~Toutanova.
\newblock {BERT}: Pre-training of deep bidirectional transformers for language understanding.
\newblock In {\em Proc.\ NAACL}, pp. 4171--4186, 2019. \href{https://doi.org/10.18653/v1/N19-1423}
{doi: {{%
10\hspace{.1pt}\discretionary{.}{%
}{.}\hspace{.4pt}18653\discretionary{/}{%
}{/}v1\discretionary{/}{%
}{/}N19\discretionary{%
}{-}{-}1423}}}


\bibitem{Devries:2017:cutout}
T.~DeVries and G.~W. Taylor.
\newblock Improved regularization of convolutional neural networks with cutout, 2017.

\bibitem{Dong:2021:HSP}
X.~Dong, A.~T. Luu, M.~Lin, S.~Yan, and H.~Zhang.
\newblock How should pre-trained language models be fine-tuned towards adversarial robustness?
\newblock In {\em Proc.\ NeurIPS}, vol.~34, pp. 4356--4369, 2021.

\bibitem{Dosovitskiy:2021:AII}
A.~Dosovitskiy, L.~Beyer, A.~Kolesnikov, D.~Weissenborn, X.~Zhai, T.~Unterthiner, M.~Dehghani, M.~Minderer, G.~Heigold, S.~Gelly, J.~Uszkoreit, and N.~Houlsby.
\newblock An image is worth 16x16 words: Transformers for image recognition at scale.
\newblock In {\em Proc.\ ICLR}, 2021.

\bibitem{Du:2022:vlp}
Y.~Du, Z.~Liu, J.~Li, and W.~X. Zhao.
\newblock A survey of vision-language pre-trained models.
\newblock In {\em Proc.\ IJCAI}, pp. 5436--5443, 7 2022. \href{https://doi.org/10.24963/ijcai.2022/762}
{doi: {{%
10\hspace{.1pt}\discretionary{.}{%
}{.}\hspace{.4pt}24963\discretionary{/}{%
}{/}ijcai\hspace{.1pt}\discretionary{.}{%
}{.}\hspace{.4pt}2022\discretionary{/}{%
}{/}762}}}


\bibitem{D’Incà:2024:fairaug}
M.~D’Incà, C.~Tzelepis, I.~Patras, and N.~Sebe.
\newblock Improving fairness using vision-language driven image augmentation.
\newblock In {\em Proc.\ WACV}, pp. 4683--4692, 2024. \href{https://doi.org/10.1109/WACV57701.2024.00463}
{doi: {{%
10\hspace{.1pt}\discretionary{.}{%
}{.}\hspace{.4pt}1109\discretionary{/}{%
}{/}WACV57701\hspace{.1pt}\discretionary{.}{%
}{.}\hspace{.4pt}2024\hspace{.1pt}\discretionary{.}{%
}{.}\hspace{.4pt}00463}}}


\bibitem{Eyuboglu:2022:domino}
S.~Eyuboglu, M.~Varma, K.~K. Saab, J.-B. Delbrouck, C.~Lee-Messer, J.~Dunnmon, J.~Zou, and C.~Re.
\newblock Domino: Discovering systematic errors with cross-modal embeddings.
\newblock In {\em Proc.\ ICLR}, 2022.

\bibitem{Fang:2023:corruption}
X.~Fang, M.~Ye, and X.~Yang.
\newblock Robust heterogeneous federated learning under data corruption.
\newblock In {\em Proc.\ ICCV}, pp. 4997--5007, 2023. \href{https://doi.org/10.1109/ICCV51070.2023.00463}
{doi: {{%
10\hspace{.1pt}\discretionary{.}{%
}{.}\hspace{.4pt}1109\discretionary{/}{%
}{/}ICCV51070\hspace{.1pt}\discretionary{.}{%
}{.}\hspace{.4pt}2023\hspace{.1pt}\discretionary{.}{%
}{.}\hspace{.4pt}00463}}}


\bibitem{Gao:2023:TBR}
Z.~Gao, K.~Huang, R.~Zhang, D.~Liu, and J.~Ma.
\newblock Towards better robustness against common corruptions for unsupervised domain adaptation.
\newblock In {\em Proc.\ ICCV}, pp. 18836--18847, 2023. \href{https://doi.org/10.1109/ICCV51070.2023.01731}
{doi: {{%
10\hspace{.1pt}\discretionary{.}{%
}{.}\hspace{.4pt}1109\discretionary{/}{%
}{/}ICCV51070\hspace{.1pt}\discretionary{.}{%
}{.}\hspace{.4pt}2023\hspace{.1pt}\discretionary{.}{%
}{.}\hspace{.4pt}01731}}}


\bibitem{Guo:2023:RTA}
Y.~Guo, D.~Stutz, and B.~Schiele.
\newblock Robustifying token attention for vision transformers.
\newblock In {\em Proc.\ ICCV}, pp. 17511--17522, 2023. \href{https://doi.org/10.1109/ICCV51070.2023.01610}
{doi: {{%
10\hspace{.1pt}\discretionary{.}{%
}{.}\hspace{.4pt}1109\discretionary{/}{%
}{/}ICCV51070\hspace{.1pt}\discretionary{.}{%
}{.}\hspace{.4pt}2023\hspace{.1pt}\discretionary{.}{%
}{.}\hspace{.4pt}01610}}}


\bibitem{Hao:2023:mixgen}
X.~Hao, Y.~Zhu, S.~Appalaraju, A.~Zhang, W.~Zhang, B.~Li, and M.~Li.
\newblock Mixgen: A new multi-modal data augmentation.
\newblock In {\em Proc.\ WACVW}, pp. 379--389, 2023. \href{https://doi.org/10.1109/WACVW58289.2023.00042}
{doi: {{%
10\hspace{.1pt}\discretionary{.}{%
}{.}\hspace{.4pt}1109\discretionary{/}{%
}{/}WACVW58289\hspace{.1pt}\discretionary{.}{%
}{.}\hspace{.4pt}2023\hspace{.1pt}\discretionary{.}{%
}{.}\hspace{.4pt}00042}}}


\bibitem{Hendrycks:2018:BNN}
D.~Hendrycks and T.~Dietterich.
\newblock Benchmarking neural network robustness to common corruptions and perturbations.
\newblock In {\em Proc.\ ICLR}, 2019.

\bibitem{Hendrycks:2021:NAE}
D.~Hendrycks, K.~Zhao, S.~Basart, J.~Steinhardt, and D.~Song.
\newblock Natural adversarial examples.
\newblock In {\em Proc.\ CVPR}, pp. 15257--15266, 2021. \href{https://doi.org/10.1109/CVPR46437.2021.01501}
{doi: {{%
10\hspace{.1pt}\discretionary{.}{%
}{.}\hspace{.4pt}1109\discretionary{/}{%
}{/}CVPR46437\hspace{.1pt}\discretionary{.}{%
}{.}\hspace{.4pt}2021\hspace{.1pt}\discretionary{.}{%
}{.}\hspace{.4pt}01501}}}


\bibitem{Hessel:2021:cider}
J.~Hessel, A.~Holtzman, M.~Forbes, R.~Le~Bras, and Y.~Choi.
\newblock {CLIPS}core: A reference-free evaluation metric for image captioning.
\newblock In {\em Proc.\ EMNLP}, pp. 7514--7528, 2021. \href{https://doi.org/10.18653/v1/2021.emnlp-main.595}
{doi: {{%
10\hspace{.1pt}\discretionary{.}{%
}{.}\hspace{.4pt}18653\discretionary{/}{%
}{/}v1\discretionary{/}{%
}{/}2021\hspace{.1pt}\discretionary{.}{%
}{.}\hspace{.4pt}emnlp\discretionary{%
}{-}{-}main\hspace{.1pt}\discretionary{.}{%
}{.}\hspace{.4pt}595}}}


\bibitem{Jaunet:2022:VXV}
T.~Jaunet, C.~Kervadec, R.~Vuillemot, G.~Antipov, M.~Baccouche, and C.~Wolf.
\newblock Visqa: X-raying vision and language reasoning in transformers.
\newblock {\em IEEE Transactions on Visualization and Computer Graphics}, 28(1):976--986, 2022. \href{https://doi.org/10.1109/TVCG.2021.3114683}
{doi: {{%
10\hspace{.1pt}\discretionary{.}{%
}{.}\hspace{.4pt}1109\discretionary{/}{%
}{/}TVCG\hspace{.1pt}\discretionary{.}{%
}{.}\hspace{.4pt}2021\hspace{.1pt}\discretionary{.}{%
}{.}\hspace{.4pt}3114683}}}


\bibitem{Kim:2025:RobustMixGen}
S.~Kim, H.~Im, W.~Lee, S.~Lee, and P.~Kang.
\newblock Robustmixgen: Data augmentation for enhancing robustness of visual–language models in the presence of distribution shift.
\newblock {\em Neurocomputing}, 619:129167, 2025. \href{https://doi.org/10.1016/j.neucom.2024.129167}
{doi: {{%
10\hspace{.1pt}\discretionary{.}{%
}{.}\hspace{.4pt}1016\discretionary{/}{%
}{/}j\hspace{.1pt}\discretionary{.}{%
}{.}\hspace{.4pt}neucom\hspace{.1pt}\discretionary{.}{%
}{.}\hspace{.4pt}2024\hspace{.1pt}\discretionary{.}{%
}{.}\hspace{.4pt}129167}}}


\bibitem{Kurakin:2017:AML}
A.~Kurakin, I.~J. Goodfellow, and S.~Bengio.
\newblock Adversarial machine learning at scale.
\newblock In {\em Proc.\ ICLR}, 2017.

\bibitem{Li:2023:blip2}
J.~Li, D.~Li, S.~Savarese, and S.~Hoi.
\newblock {BLIP}-2: Bootstrapping language-image pre-training with frozen image encoders and large language models.
\newblock In {\em Proc.\ ICML}, vol. 202, pp. 19730--19742, 23--29 Jul 2023.

\bibitem{Li:2022:BLI}
J.~Li, D.~Li, C.~Xiong, and S.~Hoi.
\newblock Blip: Bootstrapping language-image pre-training for unified vision-language understanding and generation.
\newblock In {\em Proc.\ ICML}, vol. 162, pp. 12888--12900, 2022.

\bibitem{Li:2021:AVA}
L.~Li, J.~Lei, Z.~Gan, and J.~Liu.
\newblock Adversarial vqa: A new benchmark for evaluating the robustness of vqa models.
\newblock In {\em Proc.\ ICCV}, pp. 2022--2031, 2021. \href{https://doi.org/10.1109/ICCV48922.2021.00205}
{doi: {{%
10\hspace{.1pt}\discretionary{.}{%
}{.}\hspace{.4pt}1109\discretionary{/}{%
}{/}ICCV48922\hspace{.1pt}\discretionary{.}{%
}{.}\hspace{.4pt}2021\hspace{.1pt}\discretionary{.}{%
}{.}\hspace{.4pt}00205}}}


\bibitem{Li:2021:ICA}
R.~Li, Z.~Wang, and L.~Zhang.
\newblock Image caption and medical report generation based on deep learning: a review and algorithm analysis.
\newblock In {\em Proc.\ CISAI}, pp. 373--379, 2021. \href{https://doi.org/10.1109/CISAI54367.2021.00078}
{doi: {{%
10\hspace{.1pt}\discretionary{.}{%
}{.}\hspace{.4pt}1109\discretionary{/}{%
}{/}CISAI54367\hspace{.1pt}\discretionary{.}{%
}{.}\hspace{.4pt}2021\hspace{.1pt}\discretionary{.}{%
}{.}\hspace{.4pt}00078}}}


\bibitem{Li:2020:autoaug}
Y.~Li, G.~Hu, Y.~Wang, T.~Hospedales, N.~M. Robertson, and Y.~Yang.
\newblock Dada: Differentiable automatic data augmentation.
\newblock In {\em Proc.\ ECCV}, pp. 580--595, 2020.

\bibitem{Li:2024:va4caption}
Y.~Li, J.~Wang, P.~Aboagye, C.-C.~M. Yeh, Y.~Zheng, L.~Wang, W.~Zhang, and K.-L. Ma.
\newblock Visual analytics for efficient image exploration and user-guided image captioning.
\newblock {\em IEEE Transactions on Visualization and Computer Graphics}, 30(6):2875--2887, 2024. \href{https://doi.org/10.1109/TVCG.2024.3388514}
{doi: {{%
10\hspace{.1pt}\discretionary{.}{%
}{.}\hspace{.4pt}1109\discretionary{/}{%
}{/}TVCG\hspace{.1pt}\discretionary{.}{%
}{.}\hspace{.4pt}2024\hspace{.1pt}\discretionary{.}{%
}{.}\hspace{.4pt}3388514}}}


\bibitem{Li:2023:vit}
Y.~Li, J.~Wang, X.~Dai, L.~Wang, C.-C.~M. Yeh, Y.~Zheng, W.~Zhang, and K.-L. Ma.
\newblock How does attention work in vision transformers? a visual analytics attempt.
\newblock {\em IEEE Transactions on Visualization and Computer Graphics}, 29(6):2888--2900, 2023. \href{https://doi.org/10.1109/TVCG.2023.3261935}
{doi: {{%
10\hspace{.1pt}\discretionary{.}{%
}{.}\hspace{.4pt}1109\discretionary{/}{%
}{/}TVCG\hspace{.1pt}\discretionary{.}{%
}{.}\hspace{.4pt}2023\hspace{.1pt}\discretionary{.}{%
}{.}\hspace{.4pt}3261935}}}


\bibitem{Li:2023:FACTUAL}
Z.~Li, Y.~Chai, T.~Y. Zhuo, L.~Qu, G.~Haffari, F.~Li, D.~Ji, and Q.~H. Tran.
\newblock {FACTUAL}: A benchmark for faithful and consistent textual scene graph parsing.
\newblock In {\em Proc.\ ACL}, pp. 6377--6390, July 2023. \href{https://doi.org/10.18653/v1/2023.findings-acl.398}
{doi: {{%
10\hspace{.1pt}\discretionary{.}{%
}{.}\hspace{.4pt}18653\discretionary{/}{%
}{/}v1\discretionary{/}{%
}{/}2023\hspace{.1pt}\discretionary{.}{%
}{.}\hspace{.4pt}findings\discretionary{%
}{-}{-}acl\hspace{.1pt}\discretionary{.}{%
}{.}\hspace{.4pt}398}}}


\bibitem{Lin:2004:rouge}
C.-Y. Lin.
\newblock {ROUGE}: A package for automatic evaluation of summaries.
\newblock In {\em Proc.\ ACL Workshop}, pp. 74--81, July 2004.

\bibitem{Lin:2014:MCC}
T.-Y. Lin, M.~Maire, S.~Belongie, J.~Hays, P.~Perona, D.~Ramanan, P.~Doll{\'a}r, and C.~L. Zitnick.
\newblock Microsoft coco: Common objects in context.
\newblock In {\em Proc.\ ECCV}, pp. 740--755, 2014.

\bibitem{Liu:2023:llava}
H.~Liu, C.~Li, Q.~Wu, and Y.~J. Lee.
\newblock Visual instruction tuning.
\newblock In {\em Proc.\ NeurIPS}, vol.~36, pp. 34892--34916, 2023.

\bibitem{Liu:2023:corruption}
X.~Liu, M.~Li, H.~Wang, S.~Hu, D.~Ye, H.~Jin, L.~Wu, and C.~Xiao.
\newblock Detecting backdoors during the inference stage based on corruption robustness consistency.
\newblock In {\em Proc.\ CVPR}, pp. 16363--16372, 2023. \href{https://doi.org/10.1109/CVPR52729.2023.01570}
{doi: {{%
10\hspace{.1pt}\discretionary{.}{%
}{.}\hspace{.4pt}1109\discretionary{/}{%
}{/}CVPR52729\hspace{.1pt}\discretionary{.}{%
}{.}\hspace{.4pt}2023\hspace{.1pt}\discretionary{.}{%
}{.}\hspace{.4pt}01570}}}


\bibitem{Lu:2023:SGA}
D.~Lu, Z.~Wang, T.~Wang, W.~Guan, H.~Gao, and F.~Zheng.
\newblock Set-level guidance attack: Boosting adversarial transferability of vision-language pre-training models.
\newblock In {\em Proc.\ ICCV}, pp. 102--111, 2023. \href{https://doi.org/10.1109/ICCV51070.2023.00016}
{doi: {{%
10\hspace{.1pt}\discretionary{.}{%
}{.}\hspace{.4pt}1109\discretionary{/}{%
}{/}ICCV51070\hspace{.1pt}\discretionary{.}{%
}{.}\hspace{.4pt}2023\hspace{.1pt}\discretionary{.}{%
}{.}\hspace{.4pt}00016}}}


\bibitem{Ma:2023:VAU}
J.~Ma, Y.~Bai, B.~Zhong, W.~Zhang, T.~Yao, and T.~Mei.
\newblock Visualizing and understanding patch interactions in vision transformer.
\newblock {\em IEEE Transactions on Neural Networks and Learning Systems}, pp. 1--10, 2023. \href{https://doi.org/10.1109/TNNLS.2023.3270479}
{doi: {{%
10\hspace{.1pt}\discretionary{.}{%
}{.}\hspace{.4pt}1109\discretionary{/}{%
}{/}TNNLS\hspace{.1pt}\discretionary{.}{%
}{.}\hspace{.4pt}2023\hspace{.1pt}\discretionary{.}{%
}{.}\hspace{.4pt}3270479}}}


\bibitem{Ma:2020:vulnerabilities}
Y.~Ma, T.~Xie, J.~Li, and R.~Maciejewski.
\newblock Explaining vulnerabilities to adversarial machine learning through visual analytics.
\newblock {\em IEEE Transactions on Visualization and Computer Graphics}, 26(1):1075--1085, 2020. \href{https://doi.org/10.1109/TVCG.2019.2934631}
{doi: {{%
10\hspace{.1pt}\discretionary{.}{%
}{.}\hspace{.4pt}1109\discretionary{/}{%
}{/}TVCG\hspace{.1pt}\discretionary{.}{%
}{.}\hspace{.4pt}2019\hspace{.1pt}\discretionary{.}{%
}{.}\hspace{.4pt}2934631}}}


\bibitem{Madry:2018:TDL}
A.~Madry, A.~Makelov, L.~Schmidt, D.~Tsipras, and A.~Vladu.
\newblock Towards deep learning models resistant to adversarial attacks.
\newblock In {\em Proc.\ ICLR}, 2018.

\bibitem{McInnes:2017:hdbscan}
L.~McInnes and J.~Healy.
\newblock Accelerated hierarchical density based clustering.
\newblock In {\em Proc.\ ICDMW}, pp. 33--42, 2017. \href{https://doi.org/10.1109/ICDMW.2017.12}
{doi: {{%
10\hspace{.1pt}\discretionary{.}{%
}{.}\hspace{.4pt}1109\discretionary{/}{%
}{/}ICDMW\hspace{.1pt}\discretionary{.}{%
}{.}\hspace{.4pt}2017\hspace{.1pt}\discretionary{.}{%
}{.}\hspace{.4pt}12}}}


\bibitem{McInnes:2018:UUM}
L.~McInnes, J.~Healy, N.~Saul, and L.~Grossberger.
\newblock {UMAP}: Uniform manifold approximation and projection.
\newblock {\em The Journal of Open Source Software}, 3(29):861, 2018.

\bibitem{Mintun:2021:OIB}
E.~Mintun, A.~Kirillov, and S.~Xie.
\newblock On interaction between augmentations and corruptions in natural corruption robustness.
\newblock In {\em Proc.\ NeurIPS}, vol.~34, pp. 3571--3583, 2021.

\bibitem{Papineni:2002:bleu}
K.~Papineni, S.~Roukos, T.~Ward, and W.-J. Zhu.
\newblock {B}leu: a method for automatic evaluation of machine translation.
\newblock In {\em Proc.\ ACL}, pp. 311--318, 2002. \href{https://doi.org/10.3115/1073083.1073135}
{doi: {{%
10\hspace{.1pt}\discretionary{.}{%
}{.}\hspace{.4pt}3115\discretionary{/}{%
}{/}1073083\hspace{.1pt}\discretionary{.}{%
}{.}\hspace{.4pt}1073135}}}


\bibitem{Qiu:2024:BRO}
J.~Qiu, Y.~Zhu, X.~Shi, F.~Wenzel, Z.~Tang, D.~Zhao, B.~Li, and M.~Li.
\newblock Benchmarking robustness of multimodal image-text models under distribution shift.
\newblock {\em Journal of Data-centric Machine Learning Research}, 2024.

\bibitem{Radford:2021:clip}
A.~Radford, J.~W. Kim, C.~Hallacy, A.~Ramesh, G.~Goh, S.~Agarwal, G.~Sastry, A.~Askell, P.~Mishkin, J.~Clark, G.~Krueger, and I.~Sutskever.
\newblock Learning transferable visual models from natural language supervision.
\newblock In {\em Proc.\ ICML}, vol. 139, pp. 8748--8763, 2021.

\bibitem{Radford:2019:LMA}
A.~Radford, J.~Wu, R.~Child, D.~Luan, D.~Amodei, and I.~Sutskever.
\newblock Language models are unsupervised multitask learners.
\newblock 2019.

\bibitem{Raileanu:2021:augRL}
R.~Raileanu, M.~Goldstein, D.~Yarats, I.~Kostrikov, and R.~Fergus.
\newblock Automatic data augmentation for generalization in reinforcement learning.
\newblock In {\em Proc.\ NeurIPS}, vol.~34, pp. 5402--5415, 2021.

\bibitem{Reif:2019:VAM}
E.~Reif, A.~Yuan, M.~Wattenberg, F.~B. Viegas, A.~Coenen, A.~Pearce, and B.~Kim.
\newblock Visualizing and measuring the geometry of bert.
\newblock In {\em Proc.\ NeurIPS}, vol.~32, p. 8594–8603, 2019.

\bibitem{Schiappa:2022:RAO}
M.~Schiappa, S.~Vyas, H.~Palangi, Y.~Rawat, and V.~Vineet.
\newblock Robustness analysis of video-language models against visual and language perturbations.
\newblock In {\em Proc.\ NeurIPS}, vol.~35, pp. 34405--34420, 2022.

\bibitem{Schneider:2020:IRA}
S.~Schneider, E.~Rusak, L.~Eck, O.~Bringmann, W.~Brendel, and M.~Bethge.
\newblock Improving robustness against common corruptions by covariate shift adaptation.
\newblock In {\em Proc.\ NeurIPS}, vol.~33, pp. 11539--11551, 2020.

\bibitem{Selvaraju:2017:GCV}
R.~R. Selvaraju, M.~Cogswell, A.~Das, R.~Vedantam, D.~Parikh, and D.~Batra.
\newblock Grad-cam: Visual explanations from deep networks via gradient-based localization.
\newblock In {\em Proc.\ ICCV}, pp. 618--626, 2017. \href{https://doi.org/10.1109/ICCV.2017.74}
{doi: {{%
10\hspace{.1pt}\discretionary{.}{%
}{.}\hspace{.4pt}1109\discretionary{/}{%
}{/}ICCV\hspace{.1pt}\discretionary{.}{%
}{.}\hspace{.4pt}2017\hspace{.1pt}\discretionary{.}{%
}{.}\hspace{.4pt}74}}}


\bibitem{Shao:2023:VEF}
Z.~Shao, S.~Sun, Y.~Zhao, S.~Wang, Z.~Wei, T.~Gui, C.~Turkay, and S.~Chen.
\newblock Visual explanation for open-domain question answering with bert.
\newblock {\em IEEE Transactions on Visualization and Computer Graphics}, pp. 1--18, 2023. \href{https://doi.org/10.1109/TVCG.2023.3243676}
{doi: {{%
10\hspace{.1pt}\discretionary{.}{%
}{.}\hspace{.4pt}1109\discretionary{/}{%
}{/}TVCG\hspace{.1pt}\discretionary{.}{%
}{.}\hspace{.4pt}2023\hspace{.1pt}\discretionary{.}{%
}{.}\hspace{.4pt}3243676}}}


\bibitem{Shorten:2019:augCV}
C.~Shorten and T.~M. Khoshgoftaar.
\newblock A survey on image data augmentation for deep learning.
\newblock {\em Journal of Big Data}, 6(1):1--48, 2019.

\bibitem{Shorten:2021:augNLP}
C.~Shorten, T.~M. Khoshgoftaar, and B.~Furht.
\newblock Text data augmentation for deep learning.
\newblock {\em Journal of Big Data}, 8(1):101, 2021.

\bibitem{Stefanini:2023:FST}
M.~Stefanini, M.~Cornia, L.~Baraldi, S.~Cascianelli, G.~Fiameni, and R.~Cucchiara.
\newblock From show to tell: A survey on deep learning-based image captioning.
\newblock {\em IEEE Transactions on Pattern Analysis and Machine Intelligence}, 45(1):539--559, Jan. 2023. \href{https://doi.org/10.1109/TPAMI.2022.3148210}
{doi: {{%
10\hspace{.1pt}\discretionary{.}{%
}{.}\hspace{.4pt}1109\discretionary{/}{%
}{/}TPAMI\hspace{.1pt}\discretionary{.}{%
}{.}\hspace{.4pt}2022\hspace{.1pt}\discretionary{.}{%
}{.}\hspace{.4pt}3148210}}}


\bibitem{Vaswani:2017:AIA}
A.~Vaswani, N.~Shazeer, N.~Parmar, J.~Uszkoreit, L.~Jones, A.~N. Gomez, L.~u. Kaiser, and I.~Polosukhin.
\newblock Attention is all you need.
\newblock In {\em Proc.\ NeurIPS}, vol.~30, p. 6000–6010, 2017.

\bibitem{Vedantam:2015:CCI}
R.~Vedantam, C.~L. Zitnick, and D.~Parikh.
\newblock Cider: Consensus-based image description evaluation.
\newblock In {\em Proc.\ CVPR}, pp. 4566--4575, 2015. \href{https://doi.org/10.1109/CVPR.2015.7299087}
{doi: {{%
10\hspace{.1pt}\discretionary{.}{%
}{.}\hspace{.4pt}1109\discretionary{/}{%
}{/}CVPR\hspace{.1pt}\discretionary{.}{%
}{.}\hspace{.4pt}2015\hspace{.1pt}\discretionary{.}{%
}{.}\hspace{.4pt}7299087}}}


\bibitem{Vig:2019:AMV}
J.~Vig.
\newblock A multiscale visualization of attention in the transformer model.
\newblock In {\em Proc.\ ACL: System Demonstrations}, pp. 37--42, 2019. \href{https://doi.org/10.18653/v1/P19-3007}
{doi: {{%
10\hspace{.1pt}\discretionary{.}{%
}{.}\hspace{.4pt}18653\discretionary{/}{%
}{/}v1\discretionary{/}{%
}{/}P19\discretionary{%
}{-}{-}3007}}}


\bibitem{Wang:2018:GAM}
A.~Wang, A.~Singh, J.~Michael, F.~Hill, O.~Levy, and S.~Bowman.
\newblock {GLUE}: A multi-task benchmark and analysis platform for natural language understanding.
\newblock In {\em Proc.\ EMNLP Workshop}, pp. 353--355, 2018. \href{https://doi.org/10.18653/v1/W18-5446}
{doi: {{%
10\hspace{.1pt}\discretionary{.}{%
}{.}\hspace{.4pt}18653\discretionary{/}{%
}{/}v1\discretionary{/}{%
}{/}W18\discretionary{%
}{-}{-}5446}}}


\bibitem{Zhang:2025:negaug}
Y.~Wang, D.~Gao, L.~Yi, L.~Jin, J.~Zhang, L.~Yang, and X.~Cai.
\newblock Enhancing fine-grained vision-language pretraining with negative augmented samples.
\newblock {\em Proc.\ AAAI}, 2025. \href{https://doi.org/10.1609/aaai.v36i4.20355}
{doi: {{%
10\hspace{.1pt}\discretionary{.}{%
}{.}\hspace{.4pt}1609\discretionary{/}{%
}{/}aaai\hspace{.1pt}\discretionary{.}{%
}{.}\hspace{.4pt}v36i4\hspace{.1pt}\discretionary{.}{%
}{.}\hspace{.4pt}20355}}}


\bibitem{Xuan:2025:slim}
X.~Xuan, Z.~Deng, H.-T. Lin, and K.-L. Ma.
\newblock Slim: Spuriousness mitigation with minimal human annotations.
\newblock In {\em Proc.\ ECCV}, pp. 215--231, 2024.

\bibitem{Xuan:2025:attribution}
X.~Xuan, J.~P. Ono, L.~Gou, K.-L. Ma, and L.~Ren.
\newblock Attributionscanner: A visual analytics system for model validation with metadata-free slice finding.
\newblock {\em IEEE Transactions on Visualization and Computer Graphics}, pp. 1--12, 2025. \href{https://doi.org/10.1109/TVCG.2025.3546644}
{doi: {{%
10\hspace{.1pt}\discretionary{.}{%
}{.}\hspace{.4pt}1109\discretionary{/}{%
}{/}TVCG\hspace{.1pt}\discretionary{.}{%
}{.}\hspace{.4pt}2025\hspace{.1pt}\discretionary{.}{%
}{.}\hspace{.4pt}3546644}}}


\bibitem{Xuan:2025:vista}
X.~Xuan, X.~Wang, W.~He, J.~P. Ono, L.~Gou, K.-L. Ma, and L.~Ren.
\newblock Vista: A visual analytics framework to enhance foundation model-generated data labels.
\newblock {\em IEEE Transactions on Visualization and Computer Graphics}, pp. 1--12, 2025. \href{https://doi.org/10.1109/TVCG.2025.3535896}
{doi: {{%
10\hspace{.1pt}\discretionary{.}{%
}{.}\hspace{.4pt}1109\discretionary{/}{%
}{/}TVCG\hspace{.1pt}\discretionary{.}{%
}{.}\hspace{.4pt}2025\hspace{.1pt}\discretionary{.}{%
}{.}\hspace{.4pt}3535896}}}


\bibitem{Yang:2024:labelquality}
W.~Yang, Y.~Guo, J.~Wu, Z.~Wang, L.-Z. Guo, Y.-F. Li, and S.~Liu.
\newblock Interactive reweighting for mitigating label quality issues.
\newblock {\em IEEE Transactions on Visualization and Computer Graphics}, 30(3):1837--1852, 2024. \href{https://doi.org/10.1109/TVCG.2023.3345340}
{doi: {{%
10\hspace{.1pt}\discretionary{.}{%
}{.}\hspace{.4pt}1109\discretionary{/}{%
}{/}TVCG\hspace{.1pt}\discretionary{.}{%
}{.}\hspace{.4pt}2023\hspace{.1pt}\discretionary{.}{%
}{.}\hspace{.4pt}3345340}}}


\bibitem{Yang:2024:foundation}
W.~Yang, M.~Liu, Z.~Wang, and S.~Liu.
\newblock Foundation models meet visualizations: Challenges and opportunities.
\newblock {\em Computational Visual Media}, 10(3):399--424, 2024.

\bibitem{Yeh:2024:AAG}
C.~Yeh, Y.~Chen, A.~Wu, C.~Chen, F.~Viégas, and M.~Wattenberg.
\newblock Attentionviz: A global view of transformer attention.
\newblock {\em IEEE Transactions on Visualization and Computer Graphics}, 30(1):262--272, 2024. \href{https://doi.org/10.1109/TVCG.2023.3327163}
{doi: {{%
10\hspace{.1pt}\discretionary{.}{%
}{.}\hspace{.4pt}1109\discretionary{/}{%
}{/}TVCG\hspace{.1pt}\discretionary{.}{%
}{.}\hspace{.4pt}2023\hspace{.1pt}\discretionary{.}{%
}{.}\hspace{.4pt}3327163}}}


\bibitem{Yun:2019:cutmix}
S.~Yun, D.~Han, S.~Chun, S.~J. Oh, Y.~Yoo, and J.~Choe.
\newblock Cutmix: Regularization strategy to train strong classifiers with localizable features.
\newblock In {\em Proc.\ ICCV}, pp. 6022--6031, 2019. \href{https://doi.org/10.1109/ICCV.2019.00612}
{doi: {{%
10\hspace{.1pt}\discretionary{.}{%
}{.}\hspace{.4pt}1109\discretionary{/}{%
}{/}ICCV\hspace{.1pt}\discretionary{.}{%
}{.}\hspace{.4pt}2019\hspace{.1pt}\discretionary{.}{%
}{.}\hspace{.4pt}00612}}}


\bibitem{Zhang:2018:mixup}
H.~Zhang, M.~Cisse, Y.~N. Dauphin, and D.~Lopez-Paz.
\newblock mixup: Beyond empirical risk minimization.
\newblock In {\em Proc.\ ICLR}, 2018.

\bibitem{Zhang:2022:TAA}
J.~Zhang, Q.~Yi, and J.~Sang.
\newblock Towards adversarial attack on vision-language pre-training models.
\newblock In {\em Proc.\ MM}, p. 5005–5013, 2022. \href{https://doi.org/10.1145/3503161.3547801}
{doi: {{%
10\hspace{.1pt}\discretionary{.}{%
}{.}\hspace{.4pt}1145\discretionary{/}{%
}{/}3503161\hspace{.1pt}\discretionary{.}{%
}{.}\hspace{.4pt}3547801}}}


\bibitem{Zhang:2023:sliceteller}
X.~Zhang, J.~P. Ono, H.~Song, L.~Gou, K.-L. Ma, and L.~Ren.
\newblock Sliceteller: A data slice-driven approach for machine learning model validation.
\newblock {\em IEEE Transactions on Visualization and Computer Graphics}, 29(1):842--852, 2023. \href{https://doi.org/10.1109/TVCG.2022.3209465}
{doi: {{%
10\hspace{.1pt}\discretionary{.}{%
}{.}\hspace{.4pt}1109\discretionary{/}{%
}{/}TVCG\hspace{.1pt}\discretionary{.}{%
}{.}\hspace{.4pt}2022\hspace{.1pt}\discretionary{.}{%
}{.}\hspace{.4pt}3209465}}}


\bibitem{Zhao:2023:OEA}
Y.~Zhao, T.~Pang, C.~Du, X.~Yang, C.~LI, N.-M.~M. Cheung, and M.~Lin.
\newblock On evaluating adversarial robustness of large vision-language models.
\newblock In {\em Proc.\ NeurIPS}, vol.~36, pp. 54111--54138, 2023.

\bibitem{Zhou:2023:EVP}
H.~Zhou, R.~Zhang, P.~Lai, C.~Guo, Y.~Wang, Z.~Sun, and J.~Li.
\newblock El-vit: Probing vision transformer with interactive visualization.
\newblock In {\em Proc.\ ICDMW}, pp. 118--127, 2023. \href{https://doi.org/10.1109/ICDMW60847.2023.00023}
{doi: {{%
10\hspace{.1pt}\discretionary{.}{%
}{.}\hspace{.4pt}1109\discretionary{/}{%
}{/}ICDMW60847\hspace{.1pt}\discretionary{.}{%
}{.}\hspace{.4pt}2023\hspace{.1pt}\discretionary{.}{%
}{.}\hspace{.4pt}00023}}}


\end{thebibliography}

\end{document}